\documentclass[twocolumn,twocolappendix]{aastex6}
\usepackage{mathtools}
\usepackage{natbib}
\usepackage{booktabs}
\usepackage{dcolumn}
\begin{document}


\title{X-ray morphological analysis of the Planck ESZ clusters}




\author{Lorenzo Lovisari\altaffilmark{1}}
\author{William R. Forman\altaffilmark{1}}
\author{Christine Jones\altaffilmark{1}} 
\author{Stefano Ettori\altaffilmark{2,3}}
\author{Felipe Andrade-Santos\altaffilmark{1}} 
\author{Monique Arnaud\altaffilmark{4}}
\author{Jessica D\'emocl\`es\altaffilmark{4}}
\author{Gabriel W. Pratt\altaffilmark{4}}
\author{Scott Randall\altaffilmark{1}}
\and
\author{Ralph Kraft \altaffilmark{1}}

\altaffiltext{1}{Harvard-Smithsonian Center for Astrophysics, 60 Garden Street, Cambridge, MA 02138, USA}
\altaffiltext{2}{INAF, Osservatorio Astronomico di Bologna, via Ranzani 1, 40127 Bologna, Italy}
\altaffiltext{3}{INFN, Sezione di Bologna, viale Berti Pichat 6/2, 40127 Bologna, Italy}
\altaffiltext{4}{Laboratoire AIM, IRFU/Service d'Astrophysique - CEA/DRF - CNRS - Universit\'e Paris Diderot, B\^at. 709, CEA-Saclay, F-91191 Gif-sur-Yvette Cedex, France}


\begin{abstract}
X-ray observations show that galaxy clusters have a very large range of morphologies.  The most disturbed systems which are good to study how clusters form and grow and to test physical models, may potentially complicate cosmological studies because the cluster mass determination becomes more challenging. Thus, we need to understand the cluster properties of our samples to reduce possible biases. This is complicated by the fact that different experiments may detect different cluster populations. For example, SZ selected cluster samples have been found to include a greater fraction of disturbed systems than X-ray selected samples.
In this paper we determined eight morphological parameters for the Planck Early Sunyaev-Zeldovich (ESZ) objects observed with XMM-Newton.  We found that two parameters, concentration and centroid-shift, are the best to distinguish between relaxed and disturbed systems.  For each parameter we provide the values that allow one to select the most relaxed or most disturbed objects from a sample. We found  that there is no mass dependence on the cluster dynamical state.
By comparing our results with what was obtained with REXCESS clusters, we also confirm  that indeed the ESZ clusters tend to be more disturbed, as found by previous studies.  
\end{abstract}

\keywords{X-rays: galaxies: clusters -- galaxies: clusters: intracluster medium}



\section{Introduction} \label{sec:intro}
Galaxy clusters were first detected as high concentrations of galaxies in the sky. In addition to the galaxies, there is a hot X-ray emitting intracluster medium (ICM) that accounts for the bulk of the cluster baryons. That makes X-ray surveys a powerful tool for cluster detection.  More recently,  Sunyaev-Zeldovich (SZ, \citealt{1972CoASP...4..173S}) observations have opened a new window for cluster detection and are now providing new catalogs. 
Due to the different dependence of the SZ and X-ray emission on the gas density, there is currently a debate regarding whether the two experiments are detecting the same population of galaxy clusters. In particular, since the X-ray emission scales with the square of the gas density,  X-ray surveys are more prone to detect centrally peaked, more relaxed galaxy clusters (e.g. \citealt{2011A&A...526A..79E}). Being less sensitive to the central gas density the SZ experiments detect more unrelaxed clusters (e.g. \citealt{2011A&A...536A...9P}). 

The first indications that dynamically disturbed objects are more represented in the SZ than X-ray surveys were found by \cite{2011A&A...536A...9P} and \cite{2013A&A...550A.130P} by comparing the scaled density profiles of the newly-detected SZ clusters with the ones of REXCESS clusters, a representative X-ray sample.  The SZ objects have on average flatter gas density distributions (i.e. are more disturbed morphology). 
Recently \cite{2016MNRAS.457.4515R} used the projected distance between the X-ray peak and the BCG as an indicator of relaxation for galaxy clusters. They found that X-ray selected samples tend to be more relaxed than SZ selected clusters and they interpreted the result as an indication of the cool core bias.   In a second paper, \cite{2017MNRAS.468.1917R} also investigated another morphological parameter (concentration)  and they confirmed their first paper's result and they supported the cool core bias interpretation by performing a set of simulations.   The result has been also confirmed by \cite{2017ApJ...843...76A} who compared the concentration, cuspiness and central density for the ESZ sample (only the clusters with redshift lower than 0.35) and an X-ray flux limited sample.   Although both papers obtained similar results, they provide a different explanation for the selection effects. \cite{2017ApJ...843...76A} described a simple model which predict that cool core clusters are over-represented in X-ray samples because of the Malmquist bias and were able to reproduce their results just considering the average luminosity difference between cool core and non cool core clusters. 
\cite{2017MNRAS.468.1917R}  performed a set of dedicated simulations by considering the different shape of the surface brightness profile for cool core and non cool core clusters to produce a realistic population of galaxy clusters and investigate the effect of the X-ray and SZ selection. They found that Malmquist and CC biases are probably at the origin of the different fraction of relaxed systems in the two samples but unlikely it can explain the whole difference. The above-mentioned papers are based on Planck selected samples while, interestingly, \cite{2017ApJ...841....5N},  via two other methods (i.e. centroid shifts and photon asymmetry), did not find significant differences in the observed morphology of X-ray and  SPT selected samples.  On the other hand, the different papers also used different X-ray samples (with different redshift and mass properties, as well as selection methods) for the comparison which complicate the interpretation of the different results as discussed in more detail by \cite{2017MNRAS.468.1917R}.

The different fractions of relaxed and unrelaxed systems in the samples have important implications for cosmology. In fact, the cluster mass, which is the most fundamental property to use clusters for cosmological studies, can be  over- or underestimated during a cluster merger when clusters are generally not in hydrostatic equlibrium  (e.g. \citealt{2002ApJ...577..579R}).  Moreover, compression and heating can alter both temperature and luminosity (e.g. \citealt{2001ApJ...561..621R}).  Even for relaxed systems, accurate determination of the mass requires good knowledge of both the gas density and temperature profiles up to $R_{500}$\footnote{$R_{500}$ corresponds to the radius within which the overdensity of the galaxy cluster is 500 times the critical density of the Universe.}, which are not available for many systems. And in the future almost all groups and clusters detected with eROSITA will have too few photons to measure their temperature and mass profiles (\citealt{2014A&A...567A..65B}). Thus, cosmological studies using groups and clusters of galaxies rely heavily on a detailed understanding of the scaling relations.
Measures of the dynamical states of the systems offer important information to obtain precise scaling relations and understand their scatter. For example, \cite{2015MNRAS.449..199M} showed that the identification of substructure in galaxy clusters allowed an accurate selection of relaxed systems that led  to tight constraints on the cosmological evolution of the gas mass fraction. Moreover, although to test the cosmological models we need the mass function of the whole population, the X-ray masses, obtained under the assumption of HE are more robust for relaxed clusters (e.g. \citealt{2006MNRAS.369.2013R}, \citealt{2009ApJ...705.1129L}).  A recent study by \cite{2016MNRAS.457.1522A} based on relaxed systems disfavor strong  departures from hydrostatic equilibrium and show a good agreement between X-ray and lensing masses (see also the results by \citealt{2014A&A...564A.129I}). \cite{2012ApJ...746..139A} showed that since disturbed systems tend to be less luminous and less massive they can be used in the scaling relations once the level of substructures is known and parametrized so that their positions in the mass-observable planes can be corrected (e.g. \citealt{2008ApJ...685..118V}).
 Opposed to the cosmological studies, astrophysical investigations usually focus more on highly disturbed galaxy clusters where phenomena like turbulence and particle (re-)acceleration are more prominent and easier to be investigated. Furthermore, most of these disturbed galaxy clusters are outliers in many scaling relations, thus their identification is fruitful for many different studies.

To characterize the dynamical state of a galaxy cluster requires access to a large set of information in different wavelengths (e.g. gas thermodynamical properties distribution from X-ray data; total mass distribution  from the lensing analysis) which are available only for a few individual clusters. An alternative is to compute well defined morphological parameters, making use of the relatively cheap X-ray images and profiles.  Several morphological indicators have been proposed in the last few decades: e.g., BCG-X-ray peak offset (e.g. \citealt{1984ApJ...276...38J}, \citealt{1999ApJ...511...65J}), centroid shifts \citep{1993ApJ...413..492M}, power ratios \citep{1995ApJ...452..522B, 1996ApJ...458...27B},  concentration parameter \citep{2008A&A...483...35S}, and photon asymmetry \citep{2013ApJ...779..112N}.  \cite{1996ApJS..104....1P} and \cite{2010A&A...514A..32B} showed that because of the different projection effects, none of these  methods is good in all  cases  and a combination of them might be more effective to quantify the level of substructures. 
Thus, given the large amount of new data that will be collected from current and future surveys, and that both X-ray and SZ detection methods  may have some common biases (\citealt{2012MNRAS.426.2046A}; \citealt{2015ApJ...802...34L}),  it is important to investigate the morphological indicators for large cluster samples and  {\it i)}  identify which parameter(s) most efficiently allow us to classify relaxed and disturbed objects,  and {\it ii)} verify if the X-ray  morphologies of X-ray and SZ samples are consistent.

The subdivision of relaxed and disturbed systems is itself very hard because there is no rigorous definition and a simple subdivision into two classes is probably over-simplistic.  In fact, the measure of the level of relaxation of galaxy clusters as given by the morphological parameters is a continuos function, from very relaxed systems (i.e. objects with circular X-ray isophotes and without substructures) to very disturbed objects (i.e. clear evidence of merging). While most of the systems are not on the tail of the distribution, as discussed above, several studies are only interested on the extreme systems (either the most relaxed or the most disturbed).   Thus, instead of finding parameter values that split the objects in two subsamples, for these studies, it is more important to have threshold values that selectively exclude most, if not all, the relaxed or unrelaxed systems. 

In this paper we study a set of 8 morphological parameters. Some of them are derived from the X-ray images, while others are derived from the surface brightness (SB) and  electron density profiles. The goal is to identify which are the best parameters to pinpoint the most relaxed and most disturbed galaxy clusters. 

We assume a $\Lambda$CDM cosmology with $H_0$=70 km/s/Mpc, $\Omega_{\Lambda}$=0.7 and $\Omega_m$=0.3. The outline of the paper is organized as follows. In \S2 we present the data preparation and analysis. The definitions of the morphological parameters are presented in \S3 and the results in \S4. In \S5 and \S6 we discuss the results and present our conclusions.

\section{Data analysis} 
\subsection{The sample}
Simulations have shown that the SZ quantities weakly depends on the dynamical state of the objects (e.g. \citealt{2005ApJ...623L..63M}) suggesting that SZ selected samples might be more representative of the underlying cluster population and so more appropriate for the study we are carrying out in this paper. 

ESA's Planck Mission has provided a long list of cluster candidates from which two large and statistically representative samples have been extracted: {\it i}) the ESZ sample (\citealt{2011A&A...536A...8P}), and {\it ii}) the PSZ1 cosmology sample (\citealt{2014A&A...571A..20P}). Both samples would be suitable for our analyses but we choose the ESZ sample because when we started this analyses (beginning of 2016) its XMM-Newton coverage (public data for 149\footnote{Note that since we started the data became public for another 6 clusters which are also included in the sample.} vs 142 clusters, respectively)  was larger than for the PSZ1 cosmological sample.  

The ESZ sample consists of 189 massive clusters (one is a false detection), by imposing a signal-to-noise ratio threshold of 6 on the catalogue of Sunyaev-Zel'dovich (SZ) detections above the Galactic plane ($|b|>15^{\circ}$).  The clusters span a quite broad redshift range, from 0.01 to 0.55.  XMM-Newton has observed 155 of these Planck clusters, but for five of them the observations are completely flared and cannot be used for the characterization of the cluster properties. 

\begin{figure}
\figurenum{1}
\includegraphics[width=3.2in]{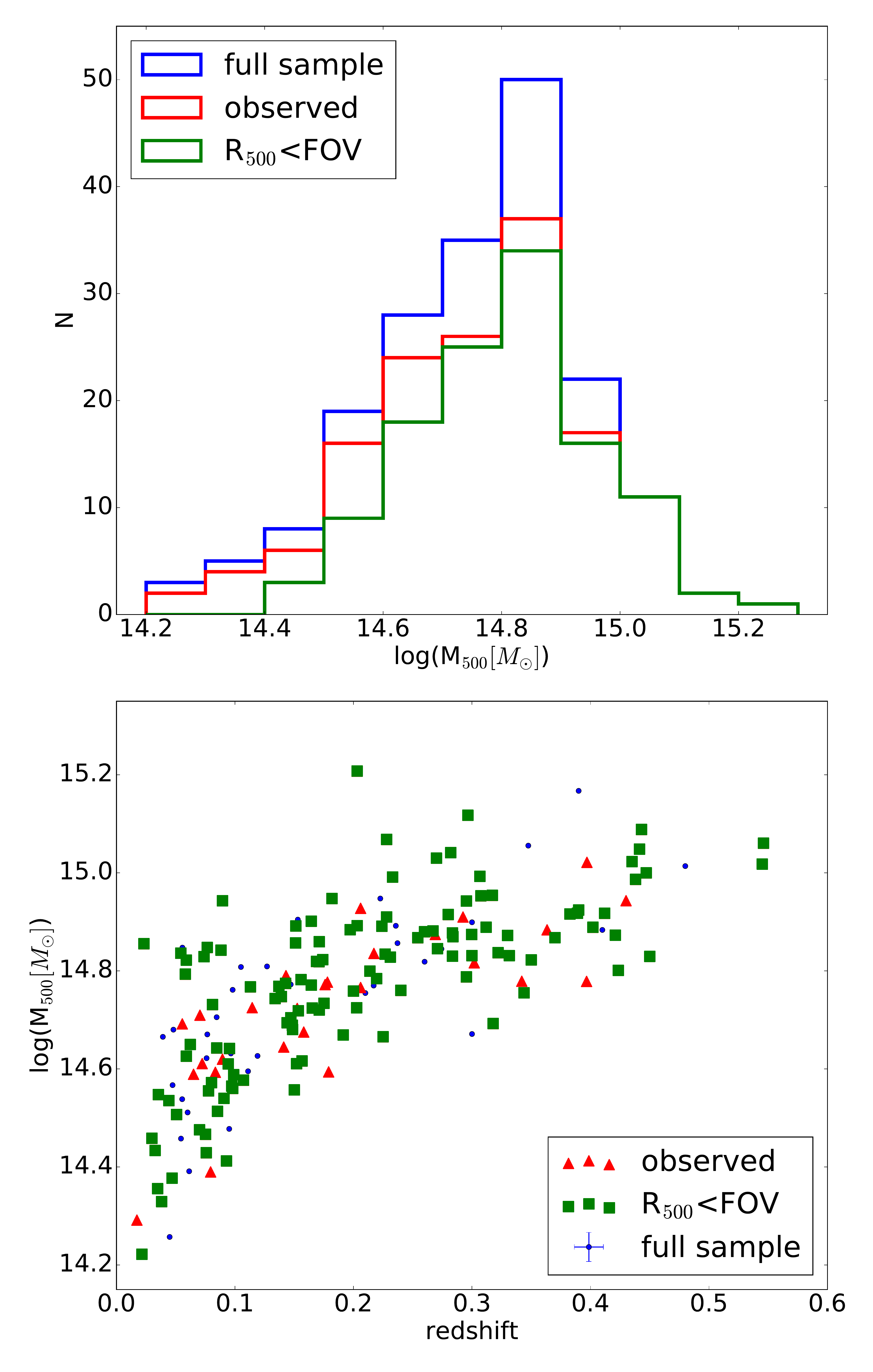}
\caption{{\it  top:} Distribution of the Planck cluster masses within $R_{500}$ for the ESZ sample.  The blue histogram refers to all the 188 objects, the red histogram to the clusters observed with XMM-Newton (excluding the flared observations), and the green histogram to the clusters that completely fit within the XMM-Newton FOV. {\it  bottom:} Mass-redshift distribution for the ESZ sample with the same colors are in the {\it top} panel. }
\end{figure}

\subsection{Data reduction} \label{sec:datared}
Observation data files (ODFs) were downloaded from the XMM-Newton archive and processed with the XMM v.16.0.0 software for data reduction. The initial data processing to generate calibrated event files from raw data was done by running the tasks {\it emchain} and  {\it epchain}. We only consider single, double, triple, and quadruple events for MOS (i.e. PATTERN$\le$12) and single for pn (i.e. PATTERN==0), and we applied the standard procedure for the removal of bright pixels and hot columns (i.e. FLAG==0) and the pn out-of-time correction.
All the data were cleaned for periods of high background due to the soft protons, following the procedure extensively described in \cite{2015A&A...573A.118L}. 

\subsection{Image analyses} \label{sec:image}
The X-ray images were created in the 0.3-2 keV energy band to maximize the signal-to-noise using a binning of 82 physical pixels corresponding to a resolution of 4.1 arcsec. 
The background subtraction was performed using a combination of blank-sky field (BSF) and filter wheel closed (FWC) observations as described in \cite{2011A&A...528A..60L}. Briefly, we selected the data sets with the most similar background properties for each cluster. We filtered the events by applying the same selection criteria used for the observations.  For each detector we added/subtracted the renormalized FWC observations to the corresponding BSF images to compensate for the difference between the out-of-field-of-view (OOFOV) events (which are a good indicator of the level of the particle background) in the observation and in the BSF data. 
The normalization factors were obtained by fitting the OOFOV events of both the observation and the BSF in the 3-10 keV energy band (for more details about this choice see \citealt{2009ApJ...699.1178Z}) with a model that includes a power-law and several gaussian lines to account for the fluorescent emission observed in both detectors. The normalization factors were derived as the ratio of the powerlaw normalization of the observation to the BSF.
The data from the three detectors were combined into a single image and divided by the combined exposure map after the MOS exposures were rescaled by a factor to account for the difference in effective area. The weighting factors have been obtained by determining the scaling of the cluster surface brightness profiles observed with each of the three detectors (see \citealt{2010A&A...514A..32B}). Thanks to this procedure all the gaps (e.g. CCD gaps) are removed from the final images. Regions exposed with less than 5$\%$ of the total exposure were  excluded\\
Point-like X-ray sources were detected with the {\it edetect-chain} task and visually inspected to discriminate between real point sources and extended cluster substructures. The latter were not removed from the data files. After removing the point sources, the holes were refilled using the CIAO task {\it dmfilth}.

\subsection{Surface brightness} \label{sec:profiles}
We determined the surface brightness (SB) profiles, centered on the X-ray peak of the main cluster component,  from the background-subtracted, vignetting-corrected images, as described in the previous section. The chosen energy band provides an optimal ratio of the source over background flux for XMM-Newton data and ensures an almost temperature-independent X-ray emission.  For the calculation of the profiles, to avoid ``humps'' in the SB profiles due to the presence of substructures or a secondary peak (e.g. in the case of an infalling system), we removed all  substructures clearly visible by eye. For bimodal mergers, the profiles were obtained independently for the two subclusters.  
The SB profiles were fitted with a double $\beta$-model:
\begin{equation}\label{eq:sb}
S_X(r)=\sum\limits_{i=1}^2 S_{0,i} \left[ 1+ \left( \frac{r}{r_{c,i}} \right)^2 \right]^{-3\beta+0.5}
\end{equation}
where $r_{c,i}$ and $S_{0,i}$ are the core radius and central surface brightness of each of the two components, respectively. This model usually provides a good fit for all the clusters once the main substructures are removed.
 
\subsection{$R_{500}$}
Galaxy clusters are at the nodes of the cosmic web, so at large radii they are expected to show signatures of accretion processes (e.g. \citealt{2006MNRAS.373.1339R}).  Hydrodynamical simulations have shown that within $R_{500}$, galaxy clusters are relatively relaxed unless a merger event modifies the existing conditions.  
Thus, $R_{500}$ represents the optimal radius for the morphological analysis to obtain a comprehensive view of the dynamical state of the clusters. Due to their low redshifts, a small fraction of objects do extend beyond the XMM-Newton FOV, so we also computed the morphological parameters within 0.5$R_{500}$.\\
Using the spectral temperature, $T_{init}$, measured in the region that maximizes the signal-to-noise in the 0.3-2 keV band, we estimated an initial $R_{500,init}$ using the equation from \cite{2005A&A...441..893A}:
\begin{equation}\label{eq:r500}
R_{500,init} = 1.104\times(T_{init}/ 5 \  {\rm keV})^{0.57}E(z)^{-1} 
\end{equation}
where E(z) is the ratio of the Hubble constant at redshift {\it z}  to its present day value. We then calculated the total gas mass ($M_{gas}$) by integrating the density profile within that radius and computed $Y_X$=$M_{gas}{\times}T_{init}$. 
By using the $M$-$Y_X$ relation given in \cite{2010A&A...517A..92A}, assuming self-similar evolution,
\begin{equation}\label{eq:MY}
E(z)^{0.4}M_{500}=10^{14.567}\left[ \frac{Y_{X,500}}{2\times10^{14}M_{\odot}keV}\right]^{0.561} M_{\odot}
\end{equation}
we estimate $M_{500}$ and $R_{500}$.  We then re-extracted a spectra in the region within  0.15-0.75$R_{500}$ to determine a new temperature and re-computed $Y_X$ using the new $R_{500}$ and $M_{500}$. The procedure is repeated until convergence, which happens when the difference between the initial and the new temperature is smaller than 1$\%$. 

The spectral analysis were done using XSPEC \citep{1996ASPC..101...17A}.
All the spectra were re-binned to ensure  at least 25 counts per bin and a minimum energy width per bin of 1/3 of the full width half maximum (FWHM) to prevent to oversample the instrument spectral resolution. Spectra were fit in the 0.3-10 keV energy range with an absorbed APEC (\citealt{2001ApJ...556L..91S}) thermal plasma with a column density from \cite{2013MNRAS.431..394W}. The EPIC spectra were fitted simultaneously, with temperatures and metallicities linked and enforcing the same normalization value for MOS, and allowing the pn normalization to vary. 

\section{Morphological parameters} \label{sec:morpar}
In the following we introduce the methods for the substructure and morphology characterization. 

\subsection{Concentration}
The concentration parameter indicates how concentrated the X-ray emission is and was first introduced by \cite{2008A&A...483...35S} as a good indicator for the presence of cooling core systems at high redshift. It is defined as the ratio of the emission within 2 different circular apertures. In this paper we use:
\begin{equation}
c=\frac{SB(<0.1R_{500})}{SB(<R_{max})}
\end{equation}  
where R$_{max}$ can be either 0.5R$_{500}$ or R$_{500}$.
While in papers using Chandra observations, it is possible to use directly the source counts estimated from the images, here we must take into account the XMM-Newton PSF. Thus, we integrated the SB profiles deconvolved with the PSF.
We discuss the relation between the concentration value obtained with and without the PSF correction in Appendix B.   

\subsection{Centroid shift}
The centroid shitft parameter is defined as the variance of the projected separation between the X-ray peak determined from the smoothed image (with a Gaussian of FWHM of 6 arcsec) and the centroid of the emission obtained within 10 apertures of increasing radius up to $R_{max}$:
\begin{equation}
w=\left[\frac{1}{N-1}\sum_i(\Delta_i-\overline{\Delta})^2 \right]^{\frac{1}{2}}\frac{1}{R_{max}}
\end{equation}
The default value for $R_{max}$ is $0.5R_{500}$, but when R$_{500}$ fits completely within the FOV we also estimated the values at $R_{max}=R_{500}$.

\subsection{Power ratios}
Introduced by \cite{1995ApJ...452..522B}, the power ratio method is motivated by the idea that the SB is a good representation of the projected mass distribution of the cluster.  The power ratios consist of a 2D multipole decomposition of the surface brightness distribution within a specified aperture and they account for the radial fluctuations where the high order moments are sensitive to smaller and smaller scales. \\
The {\it m}-order power ratio is defined as $P_m/P_0$  with \\
\begin{equation}
P_0=[a_0 \ln(R_{ap})]^2
\end{equation}
\begin{equation}
P_m=\frac{1}{2m^2R_{ap}^{2m}}(a_m^2+b_m^2)
\end{equation}
where $a_0$ is the total intensity within the aperture radius $R_{ap}$. The moments $a_m$ and $b_m$ are calculated by \\
\begin{equation}
a_m(R)=\int_{R<R_{}}S(x)R^m \cos(m\phi)d^2x
\end{equation}
and
\begin{equation}
b_m(R)=\int_{R<R_{}}S(x)R^m \sin(m\phi)d^2x
\end{equation}
where S(x) is the X-ray surface brightness at the position x=(R,$\phi$). In this paper we focus on the third (i.e. P3/P0, hereafter P30) and fourth (i.e. P4/P0, hereafter P40) moments which are sensitive to the large and small scale substructures. 

\subsection{Gini coefficient}
The Gini coefficient,  a standard economic measure of income inequality, was used for the first time in astronomy by \cite{2003ApJ...588..218A} to measure the light concentration of all galaxy types and characterize their morphology.  In this paper we use it as a measure of the X-ray flux distribution in galaxy clusters. If the total flux is equally distributed among the considered pixels, then the Gini coefficient is equal to 0, while if the total flux is concentrated into a single pixel, then its value is equal to 1. We adopt the definition from \cite{2004AJ....128..163L}:
\begin{equation}
G=\frac{1}{|\bar{K}|n(n-1)}\sum_i(2i-n-1)|{K_i}|
\end{equation} 
where $K_i$ is the pixel value in the {\it i}-th pixel of a given image, n is the total number of pixels, and $\bar{K}$ is the mean of the absolute values of all n pixels in the image. As discussed by \cite{2004AJ....128..163L}, the absolute values of $k_i$ are required because in the low surface brightness regions some pixels can result in negative values after the background subtraction. Including those negative values can yield a Gini coefficient to achieve values higher than 1. 
We note that the Gini coefficient is less sensitive to surface brightness effects and does not require a well-defined centroid (i.e. whether the flux is concentrated into a few pixels in the center or in the outer regions the obtained value is the same). That makes this parameters interesting for distant clusters and shallow observations that do not allow a precise determination of the X-ray peak. To our knowledge only \cite{2015A&A...575A.127P} computed this parameter to X-ray observations of galaxy clusters.

\subsection{Central electron density}
While the formation of a cool core in the ICM is not a fully understood process, many studies (e.g. \citealt{2010A&A...513A..37H}) have shown that the most relaxed systems tend to have a high gas density core with no significant redshift evolution (e.g. \citealt{2017arXiv170205094M}).  Under the assumption of spherical symmetry, the gas density profile can be obtained from the combination of the best-fit results from the spectral and imaging analyses as described in \cite{2015A&A...573A.118L} (see also \citealt{2010A&A...513A..37H}). 
In this paper we use the value of the density computed at 0.02$R_{500}$ to avoid the problem with the modeling close to R=0 (where the density profile may diverge) but still close enough to the cluster center to be representative of the central electron density. The central densities have not been scaled by E(z)$^-2$ but given the small redshift range the impact to the values is modest.

\subsection{Cuspiness}
Related to the density profile the cuspiness was suggested by \cite{2007hvcg.conf...48V} and it is defined as:
\begin{equation}
\alpha=-\frac{d\log\rho_g}{d\log r}
\end{equation}
where $\rho_g$ is the gas density profile and the function is computed at a fixed scaled radius of 0.04$R_{500}$.  This radius was chosen to be close enough to the core that the effect of cooling is strong but far enough to avoid the flattening of the profile due to the outflows from the central AGN. We note that although for the SB profiles we removed most of the substructures there are still a few cases where the fit is not perfectly in agreement with the data points which might bias the cuspiness for some of the clusters (in particular the most disturbed ones). 

\subsection{Ellipticity}
Although not necessarily a measure of substructures, the ellipticity is commonly defined by the ratio between the semi-minor and semi- major axis. A measure of the ellipticity can also be obtained by the power ratio P2/P0. Although we verified that the two measurements are well correlated, in this paper we used the first definition.

\section{Results}
Of the 150 analyzed clusters (excluding the ones with completely flared observations), 120 clusters have $R_{500}$ completely within the XMM-Newton field-of-view. For 28 clusters,  the estimated  $R_{500}$ extends beyond the FOV, but we could still measure their properties within 0.5$R_{500}$.
Two of the analyzed objects (AWM7 and A1060) are very nearby and therefore only a small fraction of their radius (i.e. $<$0.3$R_{500}$) lies within the FOV. Therefore, we excluded them from the analysis. 
Thus, when plotting the parameters determined within $R_{500}$, we only use 120 objects, while when we investigate the properties at $0.5R_{500}$, we make use of the full sample of 148 objects. The subsample of objects observed with XMM-Newton is representative of the full sample in terms of total masses (see {\it top} panel of Fig. 1). The same is true for the objects fitting the XMM-Newton FOV although the full sample includes a tail of low mass objects. In the {\it bottom} panel of Fig. 1 we also show the Planck mass-redshift distribution of the objects in the ESZ sample and  the XMM-Newton coverage.

\begin{figure*}[ht!]
\figurenum{2}
\includegraphics[width=7.2in]{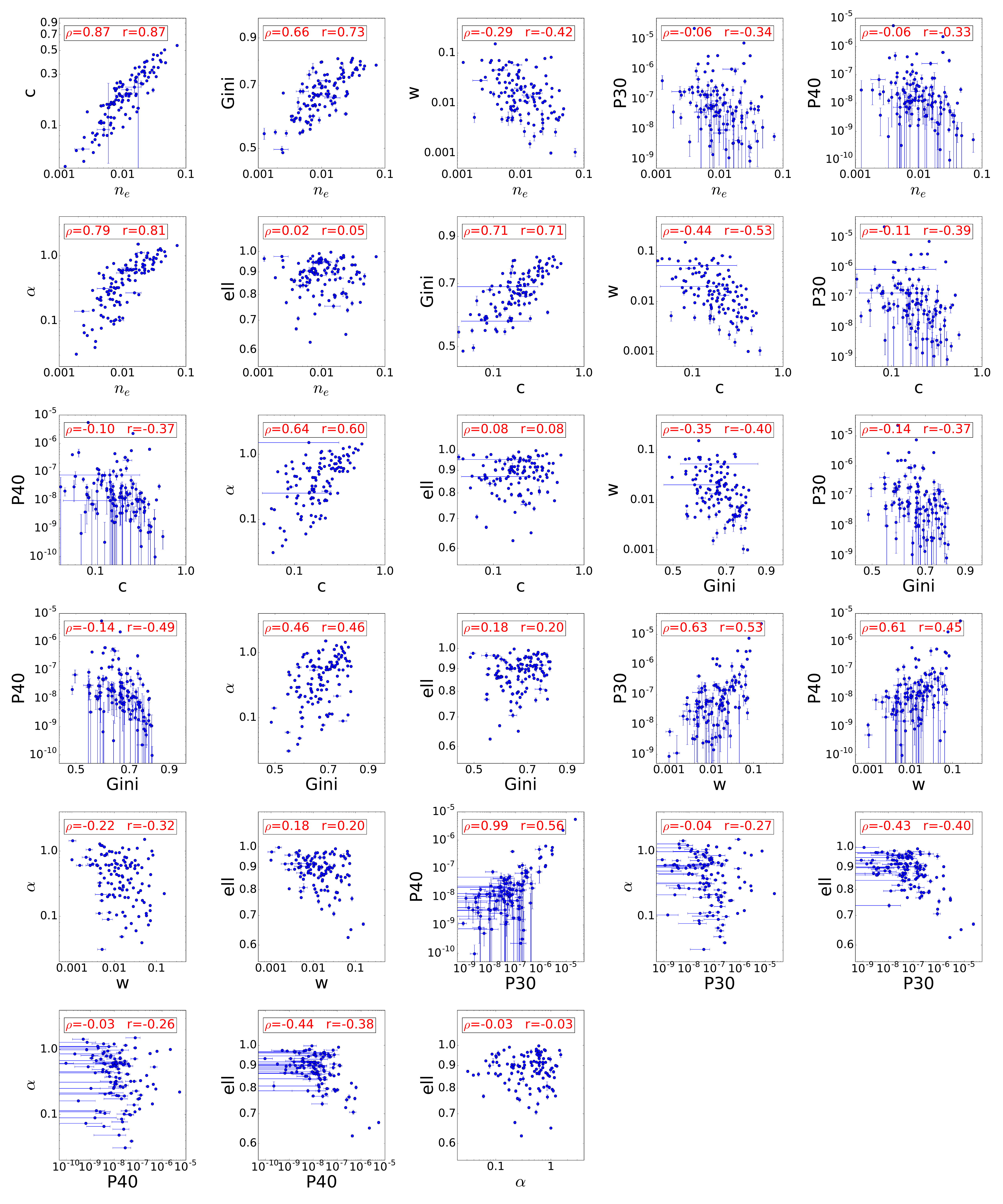}
\caption{Parameters obtained within $R_{500}$ plotted in the parameter-parameter planes. Here we show only the 120 galaxy clusters that have $R_{500}$ completely within the XMM-Newton FOV.  The $\rho$ and {\it r} values indicate the  Pearson and Spearman correlation coefficient (note that {\it r } is computed on ranks and so characterizes monotonic correlations, while $\rho$ is on true values and characterizes linear correlation). Some of the parameters show a clear and strong correlation while others are much more scattered.\\
\newline}
\end{figure*}

\subsection{Morphology parameters}
The results of the substructure analysis is summarized in Fig. 2 where we report the distribution of the parameters calculated within $R_{500}$ (see Appendix C for the parameters calculated within 0.5$R_{500}$). The uncertainties of the morphological parameters obtained  directly using the images (i.e.  Gini coefficient, centroid-shift, power-ratios, and ellipticity) have been obtained via monte-Carlo simulations as previously done by \cite{2010ApJ...721L..82C} and \cite{2016ApJ...819...36D}. For every cluster we simulated 100 versions of the X-ray images  by resampling the counts per pixel according to their Poissonian error. Similarly, to obtain the uncertainties of the parameters associated with the profiles (i.e. $n_e$, cuspiness, and concentration), we randomly varied the observational data points of the SB profiles 100 times to determine a new best fit.  Again, the randomization was driven from the Gaussian distribution with mean and standard deviations in accordance with the observed data points and the associated uncertainties.  Apart from the power ratios the uncertainties are very small (see Table H1) and they are not expected to play a big role in the correlations and classification scheme for which we used only the parameter values. Thus, the errors have been used only for illustration purposes.

\begin{figure*}[ht!]
\figurenum{3}
\includegraphics[width=7in]{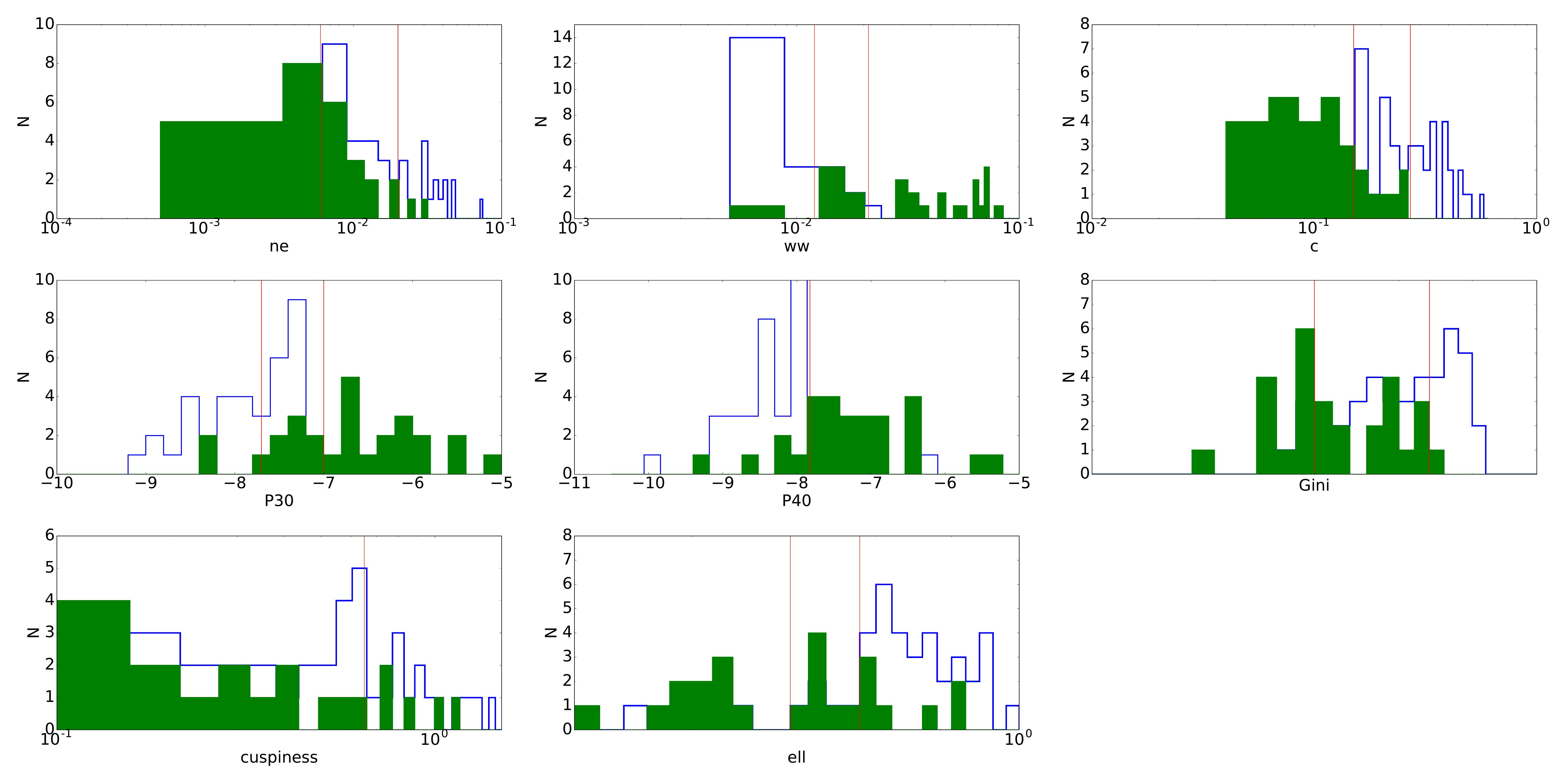}
\caption{Distribution of the relaxed (blue) and disturbed (green) systems as a function of the different parameters. The vertical red lines represent the limit values used in Tables 2 and 3.  To highlight the difference between relaxed and disturbed systems the ``mix'' objects are not plotted.} 
\end{figure*}

Although the different plots show a significant intrinsic scatter, the expected correlation between several parameters can still be observed. In fact, large centroid shifts and power-ratios, as well as small X-ray concentrations, Gini coefficients, and central densities are likely associated with disturbed clusters and so these measurements are expected to correlate with each other.    The strongest correlations (see Table I1 in the Appendix) have been obtained by comparing parameters that are more sensitive to the core properties like for example {\it c-n$_e$} ($\rho$=0.87 and r=0.87),  {\it c-Gini} ($\rho$=0.71 and r=0.71), or  {\it Gini-n$_e$} ($\rho$=0.66 and r=0.73). A good correlation is also obtained when comparing parameters that are more sensitive to the level of substructures, e.g.  {\it w-P30} ($\rho$=0.63 and r=0.53), {\it w-P40} ($\rho$=0.61 and r=0.45), or {\it P30-P40} ($\rho$=0.99 and r=0.56). A weaker correlation instead is found when comparing a parameter sensitive to the core properties with one parameter sensitive to the level of substructures (e.g. {\it  $n_e$-w} ($\rho$=-0.29 and r=-0.42), {\it  $c$-P30} ($\rho$=-0.11 and r=-0.39), or {\it  Gini-P40} ($\rho$=-0.14 and r=-0.49)).The ellipticity, while showing no correlation with the parameters sensitive to the core properties, correlates with the ones sensitive to the level of substructures.   

\subsection{Finding the most relaxed and most disturbed objects}
Each parameter has a different ability to distinguish between relaxed and unrelaxed systems. To evaluate each parameter's ability in determining the dynamical state we follow a procedure presented by \cite{2013AstRv...8a..40R} where the clusters are visually classified as relaxed, disturbed and ``mix''. A group of six astronomers inspected the images and rated the relaxation\footnote{Although the morphological disturbance (specially in 2D)  is not directly equivalent to a  departure from relaxation, as quantified for instance by the E$_{kinetic}$/E$_{thermal}$ ratio, here we refer to clusters with a low level of substructures.}  state of the clusters with a grade that ranges from 1 (most relaxed) to 4 (most disturbed). 
We then averaged the results. All the clusters with an average grade lower than 2 were classified as relaxed while the clusters with an average grade larger than 3 were classified as disturbed.  We refer to the remaining clusters with grade from 2 to 3 as ``mix''. 
Although the classification is subjective, broadly speaking, objects with circular X-ray isophotes and without substructures are classified as relaxed, double or complex objects with clear evidence of merging are classified as disturbed, and all the other with small substructures or relatively flat X-ray distribution as ``mix'' objects (see the cluster images in the Appendix).  
The distribution of relaxed and unrelaxed objects are significantly shifted with little overlap for the centroid-shift, concentration, and power ratio parameters (see Fig. 3). The overlap is larger for the central density, Gini, cuspiness and ellipticity. For these latter parameters, choosing a threshold value to classify the objects will lead either to a contaminated or incomplete sample. 
By ``contaminated'' we mean that some of the disturbed systems will be classified as relaxed (or the reverse), while for ``incomplete'', we mean that some of the relaxed objects are not recognized. If two distributions would completely shift apart, then one can choose the threshold value that allows to have a complete and not contaminated sample. However, since all the histograms overlap, one needs to find a good compromise between the completeness of the sample and its contamination. Following  \cite{2013AstRv...8a..40R}, we define two properties, the sample completeness ``C'' and the purity ``P'':
\begin{equation} 
C_r=\frac{QN(relaxed)}{TN(relaxed)} 
\end{equation}
\begin{equation} 
P_r=\frac{QN(relaxed)}{QN(relaxed+disturbed)}
\end{equation}
where QN is the number of objects above (or below) a certain threshold and TN is the total number of objects.  In a similar way we computed $C_d$ and $P_d$ for the disturbed objects. 
We also provide the purity of the sample when  ``mix'' objects are also considered: 
\begin{equation} 
P_{ext}=\frac{QN(relaxed)}{QN(relaxed+disturbed+mix)}.
\end{equation} 
In Table 1 we summarized the results of the analyses where we searched for the threshold values that optimize either C or P.  
The two cluster parameters that perform better to select the relaxed systems are the concentration and the centroid shift that, for a completeness of 100$\%$ both have  a purity of 84$\%$. The centroid shift performs better than the concentration if one searches for high purity (e.g. P$>95\%$). In fact the high purity for the concentration is reached of the higher detriment of the completeness than for the centroid-shift. The selection of the most disturbed objects results more difficult than for the relaxed objects. This is probably due to the fact that parameters that depend on models, like {\it c} or {\it $n_e$}, are better determined for relaxed than disturbed clusters. The clusters marked as disturbed but showing a rather high concentration (see Fig. D1) are usually double  or complex objects. For those clusters only the main subscluster was used for the calculation of the concentration values, slightly overestimating the concentration.\footnote{We note that A2443, A2163, and PLCKESZ124.21-36.48 still show a concentration larger than 0.15 even when using the values derived directly from the images for which the PSF effect is not taken into account.}    The centroid shift results again the best parameter to distinguish the most disturbed objects from the most relaxed but the purity of the sample is lower.

The different parameters are sensitive to different properties of the clusters. For example power-ratios and centroid shifts are sensitive to the presence of substructures, while the central density is more connected to the core properties of the clusters. Due to that some objects, which are quite relaxed and peaked in the center with some infalling substructures, can be classified differently if one use different parameters. One way  to have a more  robust selection of the most relaxed clusters in the sample is to combine more parameters. When combining two parameters, we want to keep the completeness as high as we have done with one single threshold but increase the purity of the sample. 
For example, for both concentration and centroid shift taken individually, the chosen thresholds give a 100$\%$ completeness, but ``only'' a 84$\%$ purity. Combining concentration and centroid-shift we obtain a purity of 97$\%$, while maintaining the full completeness.  In general adding a second parameter in the selection of the relaxed clusters always improves the purity of the sample although for some, the completeness drop below 90$\%$.  In Table 2 we list only the best combination of parameters. The centroid shift does a very good job in removing unrelaxed objects from the sample. In fact, the purity of all the parameters increases by 10$\%$ or more when combined with {\it w}. Also the power-ratios and ellipticity help to increase the purity of the sample although not as significantly as the centroid shift. This suggests that combining a parameter more sensitive to the level of substructures like {\it w}, {\it P30}, and {\it P40} with parameters that are more sensitive to the core properties like $n_e$ and {\it c} is the best way to identify the most relaxed clusters.  

Combining more than two morphological parameters usually reduces the completeness of the sample. For example, the only combination of three parameters that maintains the full completeness of relaxed clusters is {\it c$>$0.15}, {\it w$<$0.021}, and {\it P30$<$2E-7}.  However, that removes only very few ``mix'' objects.

\begin{table}[t!]
\tablewidth{2.5in}
\renewcommand{\thetable}{\arabic{table}}
\centering
\caption{For each parameter we indicate the limit (L) that characterize the relaxed and disturbed systems and the purity (P) and completeness (C) parameters. P$_{ext}$ refers to the purity calculated including the ``mix'' objects. For most of the parameters we provide two threshold values, one to optimize the completeness and one to optimize the purity.}
\begin{tabular}{|c|cccc|cccc|}
\hline
Par & \multicolumn{4}{c|}{relaxed} & \multicolumn{4}{c|}{disturbed} \\
 		 & L$_r$ & C$_r$ & P$_r$ & P$_{ext}$  & L$_d$ & C$_d$ & P$_d$ & P$_{ext}$ \\
\decimals
\hline
$n_e$     & $>$7.0e-3    & 0.97 & 0.74 & 0.45  & $<$3.1e-2  & 1.00 & 0.49 & 0.27  \\
$n_e$     & $>$2.5e-2      & 0.32 & 0.92 & 0.71 & $<$7e-3 & 0.54 & 0.94 & 0.39   \\
$w$     & $<$2.1e-2      & 1.00 & 0.84 & 0.48 & $>$1.2e-2  & 0.96 & 0.79 & 0.40  \\
$w$     & $<$1.2e-2      & 0.82 & 0.97  & 0.60 & $>$2.1e-2  & 0.75 & 1.00 & 0.51 \\
$c$     & $>$0.15      & 1.00 & 0.84 & 0.47 & $<$0.27  & 1.00 & 0.61 & 0.31  \\
$c$     & $>$0.27      & 0.53 & 1.00  & 0.67 & $<$0.15  & 0.75 & 1.00 & 0.60 \\
Gini     & $>$0.6      & 0.95 & 0.69  & 0.38  & $<$0.75  & 1.00 & 0.54 & 0.27 \\
Gini     & $>$0.74      & 0.45 & 0.94 & 0.68 & $<$0.60  & 0.43 & 0.86 & 0.46  \\
P30     & $<$2.0e-7      & 1.00 & 0.75 & 0.40 & $>$2.0e-8  & 0.93 & 0.57 & 0.29  \\
P30     & $<$2.0e-8      & 0.47 & 0.90 & 0.58 & $>$2.0e-7  & 0.54 & 1.00 & 0.63 \\
P40     & $<$5.0e-8      & 0.97 & 0.71  & 0.39 & $>$5.0e-9  & 0.93 & 0.58 & 0.29 \\
P40     & $<$1.0e-8      & 0.68 & 0.87  & 0.65 & $>$5.0e-8  & 0.46 & 0.93 & 0.54 \\
cusp     & $>$0.10      & 0.97 & 0.64  & 0.34 & $<$1.00  & 0.93 & 0.44 & 0.24\\
ell     & $>$0.84     & 0.97 & 0.67  & 0.40 & $<$0.95  & 0.82 & 0.40 & 0.32\\
\hline
\end{tabular}
\end{table}

\begin{table}
\tablewidth{2.5in}
\renewcommand{\thetable}{\arabic{table}}
\centering
\caption{For each parameter we indicate the limit (L) that characterize the relaxed and disturbed systems and the purity (P) and completeness (C) parameters. We list here only the combination of parameters that gives the best results. As in Table 2 P$_{ext}$ refers to the purity calculated including the ``mix'' objects.}
\begin{tabular}{|c|cc|ccc|}
\hline
Par & \multicolumn{2}{c|}{} & \multicolumn{3}{c|}{relaxed} \\
       & \multicolumn{2}{c|}{L$_r$} & C$_r$ & P$_r$ & P$_{ext}$  \\
\decimals
\hline
$c-w$  & $>$0.15  & $<$2.1e-2   & 1.00 & 0.97 & 0.59  \\
$c-P30$  & $>$0.15  & $<$2.0e-7   & 1.00 & 0.90 & 0.54  \\
$c-P40$  & $>$0.15  & $<$5.0e-8   & 0.97 & 0.93 & 0.54  \\
$c-P40$  & $>$0.15  & $<$2.0e-7   & 1.00 & 0.90 & 0.50  \\
$c-ell$  & $>$0.15  & $>$0.84   & 0.97 & 0.90 & 0.59  \\
$c-n_e$  &  $>$0.15 & $>$7.0e-3  & 0.97 & 0.86 & 0.51  \\
$n_e-w$  & $>$4.0e-3  & $<$2.1e-2   & 1.00 & 0.88 & 0.51  \\
$P30-w$  & $<$1.0e-7  & $<$2.1e-2   & 0.90 & 0.90 & 0.56  \\
$P30-P40$  & $<$1.3e-7  & $<$5.0e-8   & 0.92 & 0.74 & 0.51  \\
$Gini-c$  & $>$0.58  & $>$0.15   & 1.00 & 0.86 & 0.48  \\
$Gini-w$  & $>$0.62  & $<$2.1e-2   & 0.95 & 0.92 & 0.56  \\
$Gini-P30$  & $>$0.62  & $<$2.0e-7   & 0.95 & 0.92 & 0.48  \\
\hline
\hline
  Par    & \multicolumn{2}{c|}{} & \multicolumn{3}{c|}{disturbed} \\
       & \multicolumn{2}{c|}{L$_d$} & C$_d$ & P$_d$ & P$_{ext}$  \\
\decimals
\hline
$c-w$  & $<$0.27  & $>$1.2e-2  & 0.96 & 0.90 & 0.44  \\
$n_e-w$  & $<$3.1e-2  & $>$1.2-e2  & 0.96 & 0.84 & 0.42  \\
$P30-w$  & $>$2.0e-8  & $>$1.2e-2   & 0.89 & 0.83 & 0.42  \\
$n_e-c$  & $<$3.1e-2  & $<$0.27  & 1.00 & 0.61 & 0.31  \\
$Gini-w$  & $<$0.75  & $>$1.2e-2   & 0.97 & 0.88 & 0.49
  \\
\hline
\end{tabular}
\end{table}

\subsection{SZ vs X-ray selected clusters}
It is interesting to compare the results for the ESZ clusters with the ones obtained with the REXCESS sample, which was designed to be representative of any high quality local X-ray survey. The REXCESS clusters have been selected by their X-ray luminosity only without any specific requirement on their morphology or dynamical state.  In the cumulative plots shown in Fig. 4, we clearly see that Planck selected objects tend to be morphologically more disturbed than their X-ray counterparts.  Low p-values of the Kolmogorov-Smirnov (KS) test confirm that the two samples have indeed a different X-ray morphology if either the centroid-shift (D=0.33, p$\le$0.01), concentration (D=0.36, p$\le$0.01), or the cuspiness (D=0.31, p$\le$0.01) is used. Instead, a  p-value=0.75 confirms that the 2 distributions are indistinguishable in terms of their central densities (D=0.13).   We will give a possible explanation for this result in section 5.3.

\begin{figure*}[hp!]
\figurenum{4}
\includegraphics[width=7in]{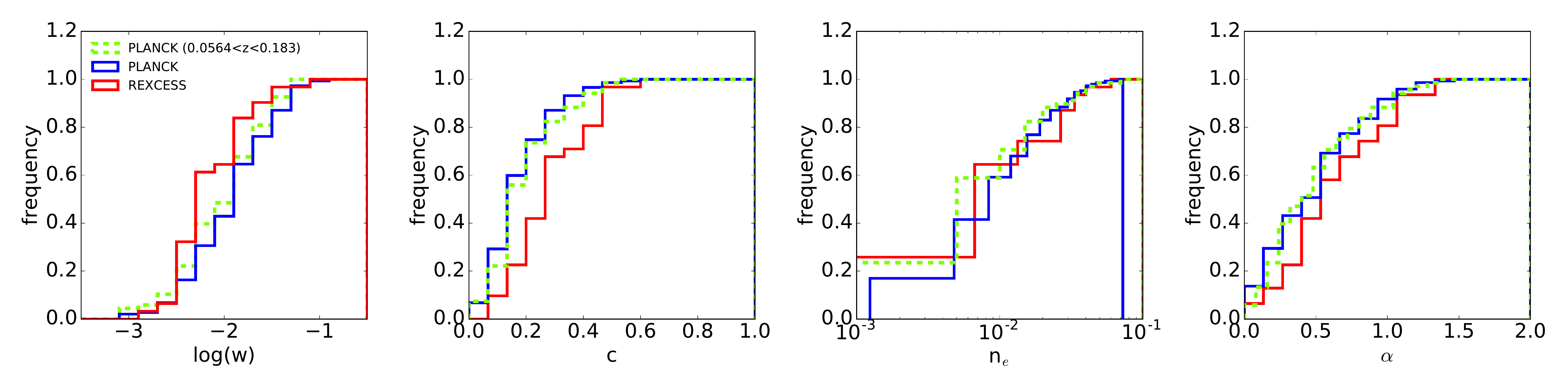}
\caption{Comparison between the centroid shift ({\it left}), concentration ({\it center-left}),  central density ({\it center-right}), and cuspiness ({\it right}) values for the ESZ (blue) and the REXCESS (red) samples calculated at $R_{500}$. The dashed green lines are the cumulative plots for the Planck clusters in the same redshift range as REXCESS. Indeed the ESZ clusters are in general more disturbed than the REXCESS clusters.}
\end{figure*}

\begin{figure*}[hp!]
\figurenum{5}
\includegraphics[width=7in,height=2.61in]{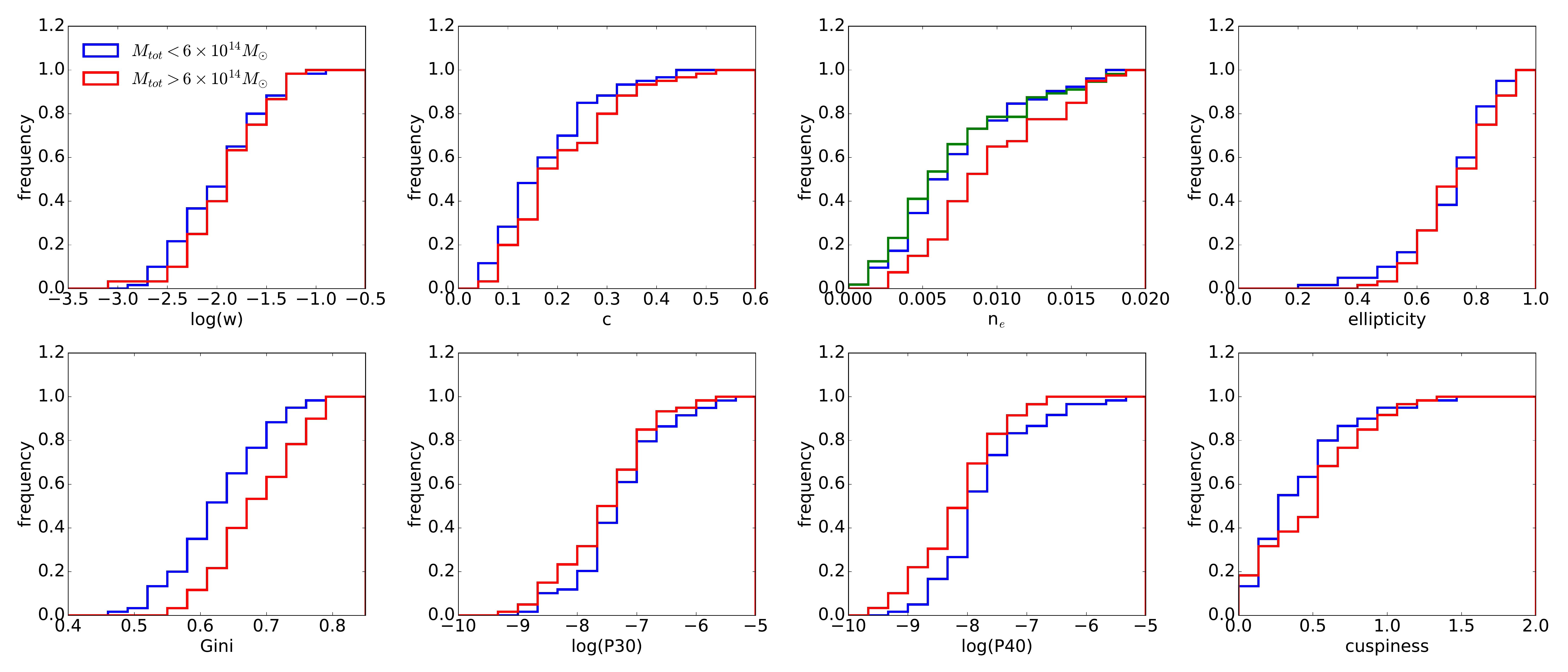}
\caption{Comparison, from top-left to bottom-right, of the  centroid shift, concentration, central density, ellipticity, Gini coefficient, P30, P40, and  cuspiness  values computed at $R_{500}$ and subdividing the sample by the total mass. \label{fig:massmor}}
\end{figure*}

\begin{figure*}[h!]
\figurenum{6}
\includegraphics[width=7in,height=2.61in]{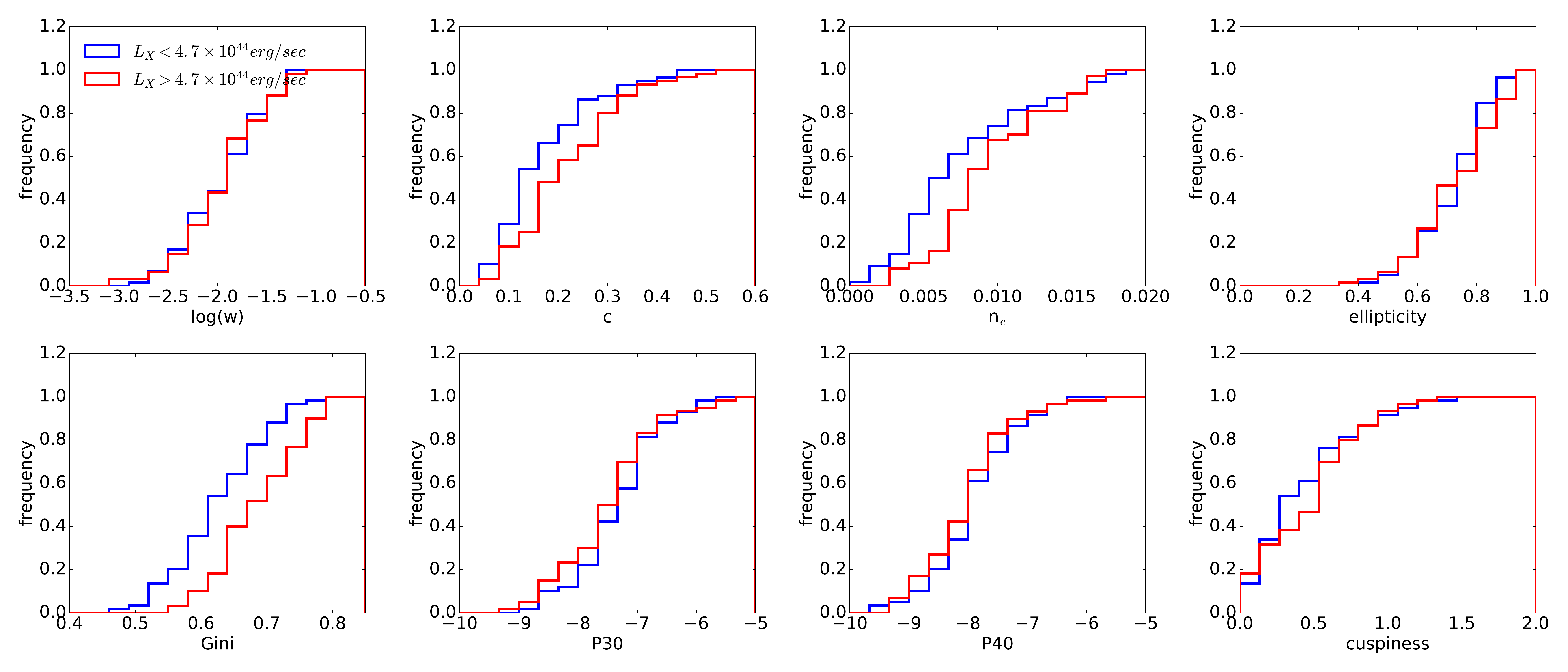}
\caption{Comparison, from top-left to bottom-right, of the  centroid shift, concentration, central density, ellipticity, Gini coefficient, P30, P40, and  cuspiness  values computed at $R_{500}$ and subdividing the sample by the total luminosity. \label{fig:luminmor}}
\end{figure*}

Since the REXCESS sample was obtained applying two redshift cuts, to have a fairer comparison we also selected from the ESZ sample only the objects in the redshift range 0.0564-0.183 (i.e. the same redshift range as the REXCESS clusters). We found no significant differences with respect to the results obtained using the full ESZ sample.  Anyway, the masses of the clusters in the ESZ sample are on average larger than those of the REXCESS sample,  although there is little to no dependence of the morphological parameter values on cluster mass (see next Section).

\subsection{Cluster properties and morphology}
We investigated the dependence of the morphological parameters on different cluster properties like total mass and X-ray luminosity. The redshift dependence of the morphological parameters will be discussed in a forthcoming paper (i.e D\'emocl\`es et al. in prep.).

We first investigated how the morphological parameters vary with total cluster mass (see Sect. 2.5 for the mass derivation), which is the most fundamental property for scaling relations of galaxy clusters. This also has some potential to impact the results on comparing X-ray and SZ samples. In fact, the cluster mass distribution for X-ray selected samples  usually extends to significantly lower masses than the SZ selected samples (e.g. \citealt{2017ApJ...843...76A}).

In Fig. 5 we show the cumulative plots for the morphological parameters computed within $R_{500}$ (see Appendix A for similar plots computed within 0.5$R_{500}$) when subdividing the sample into two mass bins to get a roughly similar number of objects in each subsample (note that the subsample with high-mass objects spans a broader range of redshift than the low-mass objects that are more peaked at low redshift). We note that a p-value $<$0.01 of the KS test confirms that the subsamples of clusters with low and high masses are significantly different. 
Although  there is a hint of a weak dependence of the centroid shift and the concentration on the cluster mass, this is not confirmed by  means of the Spearman and Pearson tests which give a relatively high probability of no correlation (see Table 3). We note that the Spearman test (which evaluates a monotonic relationship as opposed to the Pearson test that evaluates a linear relationship) predicts a very weak correlation for the concentration.  A dependence on the cluster mass is also observed with the central density, and in particular with the Gini coefficient, which shows a very strong correlation.  More massive objects have higher values of central density and Gini. This trend with the total mass disappears ({\it r}=0.10 and $\rho$=0.05) if we consider the Gini computed within 0.5$R_{500}$ instead of $R_{500}$.   

In Fig. 6 we show the cumulative plots for the same parameters when subdividing the sample by the cluster luminosities ($L_X$=4.7$\times$10$^{44}$ erg/s gives a roughly similar number of objects in each subsample). As expected we observe that the most luminous objects tend to have  higher concentrations, Gini coefficients and central gas densities. 
At the same time we do not observe any correlation between the X-ray luminosity and the centroid shift (see Table 3). Again we note that a p-value $<$0.01 of the KS test attests that the subsamples of clusters with low and  high luminosity significantly differs.

\begin{table}[t!]
\renewcommand{\thetable}{\arabic{table}}
\centering
\caption{Spearman and Pearson rank test correlation and probability for no correlation    between the cluster global properties and the morphological parameters. } \label{tab:decimal}
\begin{tabular}{|c|cccc|}
\hline
\hline
& \multicolumn{4}{c|}{$R_{500}$}  \\
Relation & \multicolumn1c{r} & \multicolumn1c{p-value}& \multicolumn1c{$\rho$} & \multicolumn{1}{c|}{p-value} \\
\hline
\decimals
M$_{500}$-c     & 0.04    & 0.69 & 0.07   & 0.43  \\
M$_{500}$-w    & 0.05    & 0.61  & 0.04    & 0.63   \\
M$_{500}$-$n_e$ & 0.27 & $<$0.01 & 0.17 & 0.07   \\
M$_{500}$-Gini     & 0.41    & $<$0.01 & 0.36  & $<$0.01 \\
M$_{500}$-cusp & 0.02 & 0.83  & -0.03  & 0.72  \\
M$_{500}$-P30  & -0.15   & 0.10   & -0.07 & 0.44  \\
M$_{500}$-P40  & -0.23   & 0.01   & -0.08 & 0.38  \\
M$_{500}$-ell  & 0.14   & 0.13   & 0.17 & 0.06  \\
\hline
L$_X$-c      & 0.28    &  $<$0.01   & 0.22   & 0.03  \\
L$_X$-w     & 0.01   & 0.91   & 0.04    & 0.68   \\
L$_X$-$n_e$ & 0.49 & $<$0.01 & 0.40 & $<$0.01  \\
L$_X$-Gini  & 0.52    &  $<$0.01   & 0.43   & $<$0.01 \\
L$_X$-cusp     & 0.20   & 0.04   & 0.17    & 0.09  \\
L$_X$-P30  & -0.09   &  0.37   & -0.04    &  0.69   \\
L$_X$-P40     & -0.03   & 0.76   & -0.18    & 0.07  \\
L$_X$-ell  & 0.11   &  0.29  & 0.07    &  0.49   \\
\hline
\end{tabular}
\end{table}

\section{Discussion}
\subsection{Morphological parameters}
In the past years several studies (e.g. \citealt{2012MNRAS.420.2120M}; \citealt{2015MNRAS.449..199M}; \citealt{2015A&A...575A.127P}) dealt with the classification of the galaxy clusters as relaxed or disturbed using different morphological parameters. Most of them are based on the analyses of Chandra data (e.g. \citealt{2015MNRAS.449..199M}; \citealt{2015A&A...575A.127P}; \citealt{2015A&A...580A..97C}) which allow a good spatial resolution but a small cluster coverage (at least in the low redshift regime where relatively short observations allow a good data quality of the images and profiles). To our knowledge the only study based on XMM-Newton data  that dealt with such classification is the one by  \cite{2010A&A...514A..32B} who performed an investigation for 31 clusters from REXCESS sample. While the unprecedented spatial resolution of Chandra allows to detect small scale substructures (but for most of the clusters we are limited by statistics) the advantage of XMM-Newton  is of course its larger FOV which allow to determine the morphological parameters up to $R_{500}$ even for relatively low redshift objects, and its large effective area which allow to collect many photons necessary to derive the morphological parameters with a great accuracy. For instance, \cite{2017A&A...598A..61B} showed that while XMM-Newton and Chandra measurements of the centroid-shift are consistent even for high redshift (z$\approx$1) massive clusters,  XMM-Newton yields $\sim$3 times smaller uncertainties than Chandra for a given exposure time.

We presented a set of eight morphological parameters to constrain the dynamical state of the ESZ  galaxy clusters. Three of them (i.e. centroid-shift {\it w}, and two power-ratios P30 and P40) are sensitive to the presence of substructures, which are certainly indicative of a dynamically active system. Since the absence of substructures does not necessary imply that the cluster is relaxed we complemented the results with the concentration, central density, cuspiness, and Gini coefficient that instead are more sensitive to the core properties (i.e. how peaked the gas distribution is, which is an indication of relaxation) of the systems. Finally, an old merger where most of the substructures have been washed out can still be identify by the strong elliptical shape of the cluster. 



Some of these parameters show a very strong correlation with each other (e.g. $n_e$-c or P30-P40) while other show no correlation (e.g. P30-cuspiness or Gini-ellipticity).  This is true both within 0.5$R_{500}$ or considering the full volume within $R_{500}$.  
 The correlation is, not surprisingly, tighter when both the considered parameters are more sensitive to the core properties ($n_e$, Gini, and c) or, with somehow weaker correlation coefficients, when both considered parameters are more sensitive to the level of the substructures (e.g. w, P30). Instead, when two parameters sensitive to different features (i.e. one to the core properties and one to the presence of substructures) are considered, the correlation is much weaker.
This is most probably due to the fact that some clusters with infalling substructures still host a cool core in the center. Thus, on one side the concentration, Gini, and central density suggest a more relaxed object, on the other, {\it  w}, P30, and P40 parameters which are more sensitive to the level of substructures shows a more disturbed dynamical state  for the cluster.  
Nevertheless almost all the correlations are quite scattered, which suggests that using one parameter alone to classify relaxed or unrelaxed clusters can yield to misleading results. 

The concentration, centroid-shift, and, at a lower level the power-ratios are able to  separate the distribution of relaxed and unrelaxed systems with very little overlap (see histograms in Fig. 3). The other parameters (i.e. cuspiness, Gini, n$_e$, and ellipticity) instead show a large number of objects in the overlap region, making these less powerful to distinguish between different dynamical states.  All the objects visually identified as relaxed have a concentration higher than 0.15 and centroid shift smaller than 0.021.  Using only one of these threshold values allows us to build a subsample of objects that include all the relaxed  systems with a 16$\%$ contamination of disturbed systems  which make them the best parameters to characterize the cluster dynamical state among the ones investigated in this paper.  Indeed, the contamination is higher if the ``mix'' objects are included in the calculation. On the other hand, by definition,  these objects do not show clear merging features and they might also be relaxed.   

The Gini coefficient strongly correlates with the concentration. This makes it very attractive, and in theory very powerful, because contrary to the concentration it is insensitive to the choice of the X-ray center. Unfortunately, our results show that Gini is not as efficient as the concentration and/or the centroid shift  to distinguish the most relaxed objects from the most disturbed. One of the reasons is that double or complex objects (e.g. A2744 or PLCKESZ266.02-21.25), for which we should expect low Gini coefficients, show prominent substructures where the flux is concentrated which effect is to increase the Gini values. On the contrary, some relaxed clusters (e.g. A2175 or A2426) have surprisingly a low Gini coefficient, probably due to the Gini definition used in this paper.  In fact, as discussed in Sect. 3.4 we assigned positive value to the pixels that scattered below the sky level. As explained by \cite{2004AJ....128..163L} this correction is not able to recover the ``true'' Gini coefficient for images with low S/N. Thus, for galaxy clusters with relatively shallow observations and with very few photons in the outer regions the Gini value can be underestimated. Despite that, by combining it with {\it w} which easily identifies the most relaxed objects, {\it c}, or {\it P30}, we can obtain a cleaner sample of relaxed (disturbed) systems. 

\subsection{Most relaxed and most disturbed systems}
Current and future surveys will provide us with very large galaxy cluster catalogs making eventual visual classification difficult as well as prone to the observer-bias problems connected with that. Thus, we searched the best combination of parameters that will allow us to robustly identify the most relaxed and most disturbed systems in a sample. Naively, one aims to detect all, and only, the relaxed or disturbed objects. We showed that combining the concentration and centroid-shift values allow us to obtain all the relaxed (disturbed) objects with a very small contamination by merging (relaxed) systems.  The power of this combination was also demonstrated by \cite{2010ApJ...721L..82C} who used those parameters to show the relation between galaxy cluster mergers and the presence of extended radio halos.

Following \cite{2013AstRv...8a..40R}  we also define a new general parameter M as a combination of the concentration and the centroid shift. That allows to have one single value to classify the X-ray morphology, and to distinguish between the relaxed and disturbed systems. The definition of the parameter is the following:
\begin{equation}
M=\frac{1}{2}\left(\frac{c-0.15}{c_{quar}-c_m}-\frac{w-0.021}{w_{quar}-w_m}\right)
\end{equation}
where $c_m$ and $w_m$ are the medians of the concentration and the centroid shift, respectively, and $c_{quar}$ and $w_{quar}$ are the first or the third quartile depending if the parameter of the specific cluster is smaller or larger than its median (see \citealt{2013AstRv...8a..40R}  for more details). In Fig. 7 we show the distribution of the M values for all the clusters. The distribution clearly shift apart for the relaxed and disturbed objects. All the relaxed objects have an M value greater than 0.5 while, apart from one cluster, all the disturbed objects have an M value lower than 0.5.

\begin{figure}[t!]
\figurenum{7}
\includegraphics[width=3.5in]{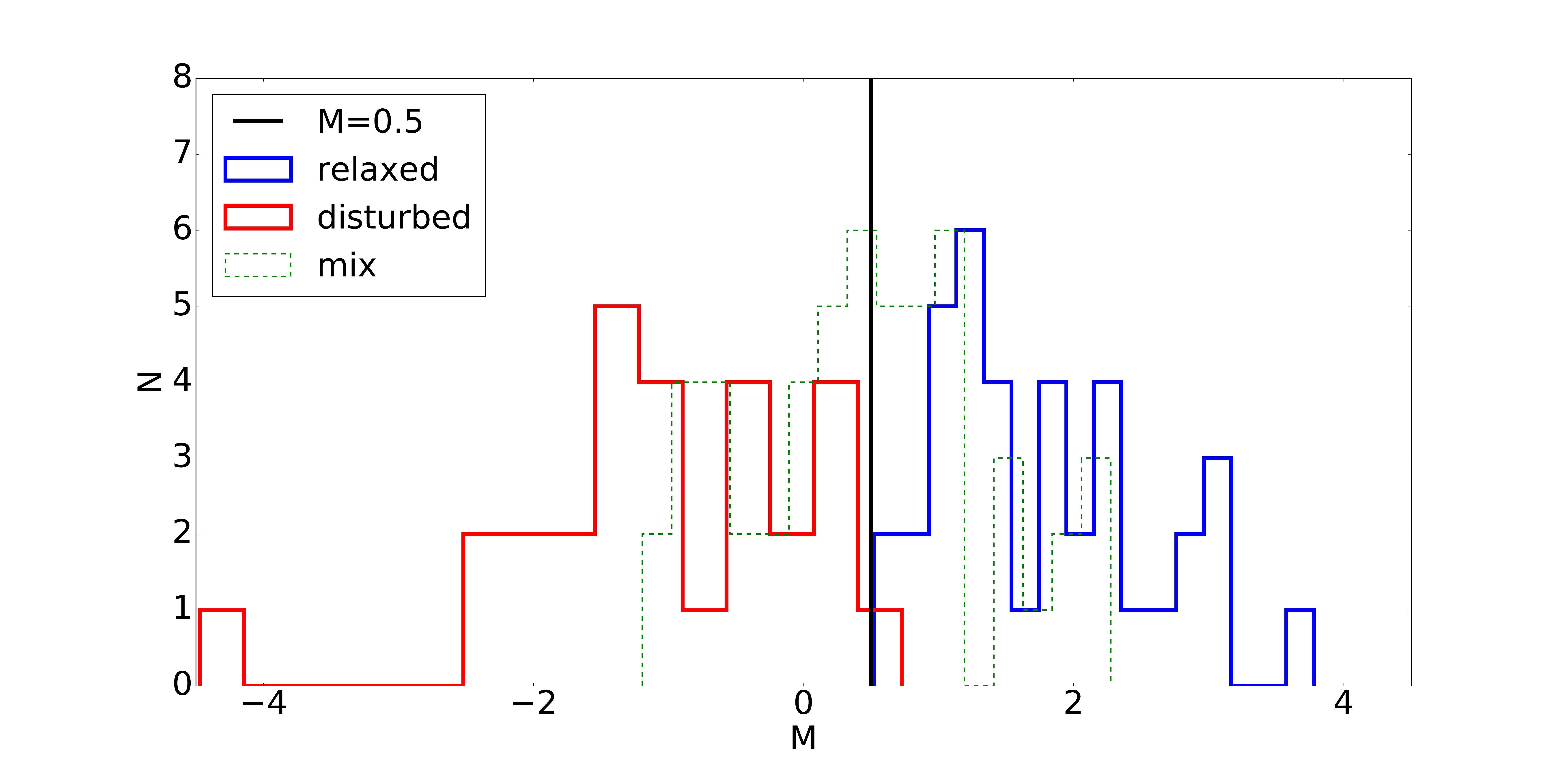}
\caption{Distributions of the M parameter as defined in Eq. 13. Blue and Red histograms refer to visually classified  relaxed and disturbed systems while the dashed green distribution refer to the ``mix''  objects.}
\end{figure}

Alternatively to the centroid-shift, one can use the power-ratios, the ellipticity, or the central electron density although their ability to remove the most disturbed or relaxed systems is lower than {\it w}. 
While in general combining more parameters  always improves the purity of the sample, the best results are obtained when parameters more sensitive to identify substructures are combined with parameters more sensitive to the core properties. This is because some clusters, for which the core is still prominent (i.e. tend to give high concentration), on large scales show the presence of substructures (e.g. infalling systems) that are indicative that the system is not fully relaxed. Opposite to that,  the bulk of the emission of  some clusters appears to be relaxed on large scales (i.e. no presence of X-ray substructures) but at the same time they do not show a bright and peaked core suggesting that either an earlier merger prevented its formation or the merger is ongoing along the line of sight  to which morphological parameters such as {\it w} are not very sensitive. Getting rid of these systems assures that only the most relaxed objects are included in the sample.  

While our results suggest that the centroid-shift, eventually paired with a second parameter,  is the most  powerful parameter in distinguishing relaxed and disturbed systems, \cite{2015A&A...575A.127P} suggested that unlike parameters sensitive to the core properties, parameters more sensitive to the substructures (like {\it w}) are not able to efficiently classify the galaxy clusters dynamical state. In particular, they investigate the smoothness and asymmetry parameters that \cite{2013AstRv...8a..40R} found very promising with simulations. \cite{2015A&A...575A.127P} noted that the values of these two parameters depends on the cluster exposure time and S/N which is also true for the power ratios but not for the centroid shift (at least in the count regime investigated in this paper) as shown in the Appendix A. So, while it is true that the ability of some parameters to distinguish different dynamical states indeed depends on the quality of the data, our results indicate that the centroid-shift works well also in a relatively low-count regime.

The combination of different parameters to identify the most relaxed galaxy clusters from a sample was also done by \cite{2015MNRAS.449..199M}, who introduced the symmetry-peakiness-alignment criterion. Their strategy was based on parameters which do not need a complete imaging coverage. However, they show that their parameters strongly (anti-)correlate with both concentration and centroid-shift suggesting that they are able to measure the same X-ray features. We note that although they use different parameters they also combine a parameter more sensitive to the core properties (i.e. the peakiness) with parameters more dependent to the large scale inhomogeneities (i.e. symmetry and alignment) similarly to what done in this paper. We note that they restricted their analysis to relatively small radii which may impact the final results (e.g. see the comparison between the parameters computed at 0.5R$_{500}$ and R$_{500}$ in Appendix C) and their analysis was optimized to find the most relaxed systems while here we also provide threshold values for selecting the most disturbed galaxy clusters.

\subsection{Dependence on the cluster properties}
The hierarchical structure formation model predicts that massive clusters form through episodic mergers of small mass units. Because of that,  statistically  one might expect to find the most massive objects in a more disturbed dynamical state.  Our morphological analysis supports this scenario only if we use the central density or the Gini coefficient  as the reference metric.  The dependence of the central density on the cluster mass is due to the fact that the more massive clusters show a higher gas fraction (e.g. \citealt{2009A&A...498..361P}; \citealt{2015A&A...573A.118L}). The centroid shift and concentration parameters instead show no or  very weak dependence on the total mass. The correlation of the mass with Gini is probably an artifact due to the change of the negative pixels in positive pixels required to avoid Gini values higher than 1 that would complicate the interpretation of the results. As discussed in the previous section (see also Appendix A) using the absolute values may bias down the result of the Gini coefficient when dealing with low S/B regions. Thus, for low mass clusters (and so less luminous) the outer regions have a lower S/N (for a given exposure) and the Gini values will be biased down. When the outer regions are removed (e.g. when the parameters are computed within 0.5R$_{500}$) from the calculation, and so most of the negative pixels are not included, the correlation between the total mass and Gini disappear (see Table D2).  

The mass dependence on the morphological parameters  (centroid shift and power ratios) was also investigated by \cite{2010A&A...514A..32B}. Their results also show no dependence of the morphological parameters on cluster mass. \cite{2017MNRAS.468.1917R} used the concentration parameter (but defined as the counts ratio within 40 and 400 kpc) to estimate the CC-fraction for the Planck cosmology sample, subdividing the objects into two subsamples with different masses. They found hints of a higher CC fraction for the most massive objects but at low significance. Hint of a higher CC fraction was also found by \cite{2015MNRAS.449..199M}.  
\cite{2017arXiv170205094M} did not find any evidence of a redshift evolution in the fraction of merging clusters which is consistent with an evolving merger rate (as predicted by the simulations of  \citealt{2010MNRAS.406.2267F}), provided that the relaxation time scale also evolves with redshift.  Simulations have also shown that the merger rate has a weak dependence on halo mass (e.g. \citealt{2010MNRAS.406.2267F}) but we argue that an infalling subhalo of a given mass would have a larger impact to a low mass cluster than to a massive system (e.g. it will take longer to restore the equilibrium). The fact that we do not observe major differences in the fraction of disturbed systems between low and high mass clusters support the scenario suggested  by \cite{2017arXiv170205094M} who proposed that halos  assemble rapidly at high redshift and then the growth get slower. We note that our clusters do no span a very large range of masses, with 85$\%$ of them in the range 3-9$\times10^{14}M_{\odot}$. So, the analysis should be extended to a larger range to confirm the current results.

Given the L$_X$-M relation, it is not surprisingly that there is also no correlation between the X-ray luminosity and the centroid-shift, and power-ratios. There is instead a weak correlation with the concentration. As noted by \cite{2010A&A...514A..32B}, this is probably due to a selection effect. In fact, for a given mass, CC clusters have in general higher X-ray luminosities than non-CC clusters (\citealt{2009A&A...498..361P}).  Since, these CC clusters are usually more relaxed, they lie on the high luminosity side.

As for the mass, we found that both central density and Gini are correlated also with the luminosity, which is expected given the L$_X$-M relation. The luminosity shows also a dependence with the concentration with the most luminous having as expected a higher concentration in the center.  Thus, indeed the luminosity is well correlated with parameters more sensitive to the core properties. It does not show instead any dependence with the parameters more connected with the presence of substructures. If we assume that the state of relaxation  of a cluster is connected with the presence and number of substructures then the  X-ray global properties  are not helpful to determine its dynamical state. 
A similar result was obtained with the REXCESS sample by \cite{2010A&A...514A..32B}. 

\subsection{ESZ vs REXCESS}
We compared different morphological parameters with the ones derived for the REXCESS sample.  Apart from the central density, all of them confirm that Planck selected objects are generally more disturbed than the X-ray selected sample. Since the two samples have a different redshift distribution and given the weak dependence on the redshift for some parameters, we also constrained the analyses to the same redshift range. However the difference between the two samples remained. The two samples also have a different mass distribution, but as previously discussed, only the central density and Gini show a mass dependence, so that should not have impact on the comparison. The mass dependence of the central density is likely the explanation for the lack of difference between the ESZ and REXCESS values. In fact, ESZ clusters are more massive than REXCESS clusters, so they have in general higher central density (M$_{tot}-n_e$ relation), but that is counterbalanced by the fact that the ESZ clusters are morphologically more disturbed and have a lower central density (e.g. $n_e-w$, $n_e-P_{30}$). This implies that if one uses the central density to compare the properties of SZ and X-ray selected samples one must assure that the samples span a similar range in redshift or total masses.  

Recently, \cite{2017MNRAS.468.1917R} compared the fraction of cool-core clusters in the Planck cosmological sample and  the X-ray selected sample MACS (\citealt{2012MNRAS.420.2120M}) using the concentration parameter as a proxy for CC. They found that the CC fraction in the X-ray selected sample is higher than in Planck.  We note that their definition of {\it c} is taken from the original work of \cite{2008A&A...483...35S} (i.e.  c=SB(r$<$40kpc)/SB(r$<$400kpc)) and so is different from the one used in this paper.  

The use of the different definition for the concentration (i.e. using physical radii or a fixed fraction of $R_{500}$) might have impact on the determination of the specific fraction of CC.
In fact, \cite{2017arXiv170205094M} analyzing the density profiles of a large sample of clusters suggested that the size of CC are not evolving with time.  Accordingly with that it would make more sense to use physical instead of scaled radii to compute the concentration. On the other hand,  \cite{2010A&A...513A..37H} analyzing the temperature profiles of the HIFLUGCS sample (see \citealt{2002ApJ...567..716R} for more details) showed that the hotter (and so more massive) clusters have larger core radii (in physical scale) but in general smaller than 0.1$R_{500}$ (used for example in this paper). Thus, while the size of the cores is not evolving, it is plausible that clusters of different masses host cores with different sizes making both approaches suitable for these kind of investigations. \cite{2017ApJ...843...76A} compare the CC fraction in SZ and X-ray samples using both physical radii or a fixed fraction of $R_{500}$. Both methods  point toward a larger number of relaxed objects in the X-ray selected samples although the two approaches gives a different fraction of CC.  The different fraction may depend on the threshold values used to classify a cluster as CC but also from the fact that their sample span a broader range of masses. 

Other studies (e.g. \citealt{2015MNRAS.449..199M}, \citealt{2017MNRAS.468.1917R}, \citealt{2017ApJ...843...76A}) have shown that Planck selected clusters tend to be morphologically more disturbed than they X-ray counterpart in agreement with the fact that the selection of X-ray cluster samples is significantly biased towards cool-core clusters (e.g. \citealt{2011A&A...526A..79E};  \citealt{2017MNRAS.468.1917R}; \citealt{2017ApJ...843...76A}). The recent papers from \citealt{2017MNRAS.468.1917R} and \citealt{2017ApJ...843...76A}  computed several morphological parameters sensitive to the core properties (i.e. {\it c, $n_e$},  and {\it cuspiness}), and although a direct comparison is not possible due to the different definition of the parameters and of the used $R_{500}$ we note that  our results qualitatively agree with their finding of a larger fraction of relaxed systems in the X-ray selected samples. However, contrary to what found by   \citealt{2017ApJ...843...76A}  we did not find any significant difference in the central electron density of the Planck and X-ray selected clusters.
Among the different parameters computed in \cite{2017ApJ...843...76A} the central electron density is the one showing the smallest difference in terms of CC fraction between SZ and X-ray selected clusters. As we showed more massive clusters have a higher electron density so the comparison between the different samples depends on the relative mass distribution difference. Moreover, we note that \citealt{2017ApJ...843...76A} computed the central electron density at 0.01$R_{500}$ while our values were computed at 0.02$R_{500}$. Since the central density value is model dependent (i.e. the way one extrapolates to the center) the choice of R$_{500}$ at which to compute $n_e$ can partially explain the different results.

Previous studies dealing only with X-ray data (e.g. \citealt{2017MNRAS.468.1917R}, \citealt{2017ApJ...843...76A}) used parameters sensitive to the core properties to investigate the differences between X-ray and Planck selected samples. In this paper,  we confirmed that Planck selected clusters tend to be morphologically more disturbed than they X-ray counterpart by using the centroid-shift which is more related to the dynamical state of the clusters rather their core properties.  This result is in disagreement with what obtained by \cite{2017ApJ...841....5N} who did not find any morphological difference between an SPT sample and 400d, an X-ray selected sample (see \citealt{2007ApJS..172..561B} for more details). Indeed the SPT sample is at z$>$0.4, while the ESZ is at z$<$0.55 (and mostly at much lower redshift), so maybe the difference is more important at low z.
However, that raises the question of what is the origin of the difference between Planck and X-ray samples and/or why the SPT clusters do not show the same morphological differences. That will require a dedicated paper for which the parameters are estimated consistently (e.g. same definition and same algorithm for the parameter calculation) for well defined and comparable samples (e.g. similar mass and redshift distribution).

\section{Summary and Conclusion} 
In this paper we investigated several morphological parameters for the ESZ sample to identify which parameters are more powerful to pinpoint the most relaxed objects from the sample. We also investigated whether the occurrence of substructures or the presence of cool cores depends on  the cluster properties, incuding L$_X$ and  total mass. Finally by comparing our results with what has been obtained with REXCESS data, a representative X-ray selected sample, we investigated if the SZ and X-ray surveys are selecting the same population of galaxy clusters. Our main conclusions are the following:
\begin{itemize}
\item[-] Concentration and centroid-shift are the parameters that perform better  in identifying relaxed  systems. All the objects visually classified as relaxed have a concentration higher than 0.15 and a centroid-shift value lower than 2.1E-3.   
\item[-] Identifying  the most disturbed systems by using the morphological parameters is in general more difficult than identifying the most relaxed ones. 
\item[-]  Combining two parameters is a more efficient way to select a complete and pure sample of relaxed or disturbed systems. In particular, it is best to combine parameters more sensitive to substructures (e.g. {\it w}, {\it P30}, and {\it P40}) with parameters more sensitive to the core properties (e.g. {\it n$_e$} and {\it c}).  The best results are obtained by combining the concentration with the centroid-shift.
\item[-] Apart from the central gas density and Gini coefficient, there is no dependence on the morphological parameters with the total cluster mass. The  $M_{tot}-n_e$ correlation implies that the central density can be used to compare different samples only if they span the same mass range.\\
\item[-] Samples of SZ selected clusters tend to be more dynamically  disturbed (i.e. high centroid shift and low concentration and central density) than the X-ray selected samples in agreement  with what has been found by other recent studies.
\end{itemize}

\acknowledgments
We thank the anonymous referee for the useful report which helped to improve the quality of the paper. We acknowledge Gerrit Schellenberger and Reinout van Weeren for helping with the visual classification of the clusters. L.L. acknowledges support from the Chandra X-ray Observatory grant GO3-14131X and by the Chandra X-ray Center through NASA contract NNX16AH31G. 
C.J. and W.R.F. are supported by the Smithsonian Institution.
F.A-S. acknowledges support from {\em Chandra} grant GO3-14131X.
M.A., G.W.P., and J.D. acknowledge support from the European Research Council under the European Union's Seventh Framework Programme (FP7/2007-2013) / ERC grant agreement no. 340519.

\appendix
\begin{figure*}
\figurenum{A1}
\plottwo{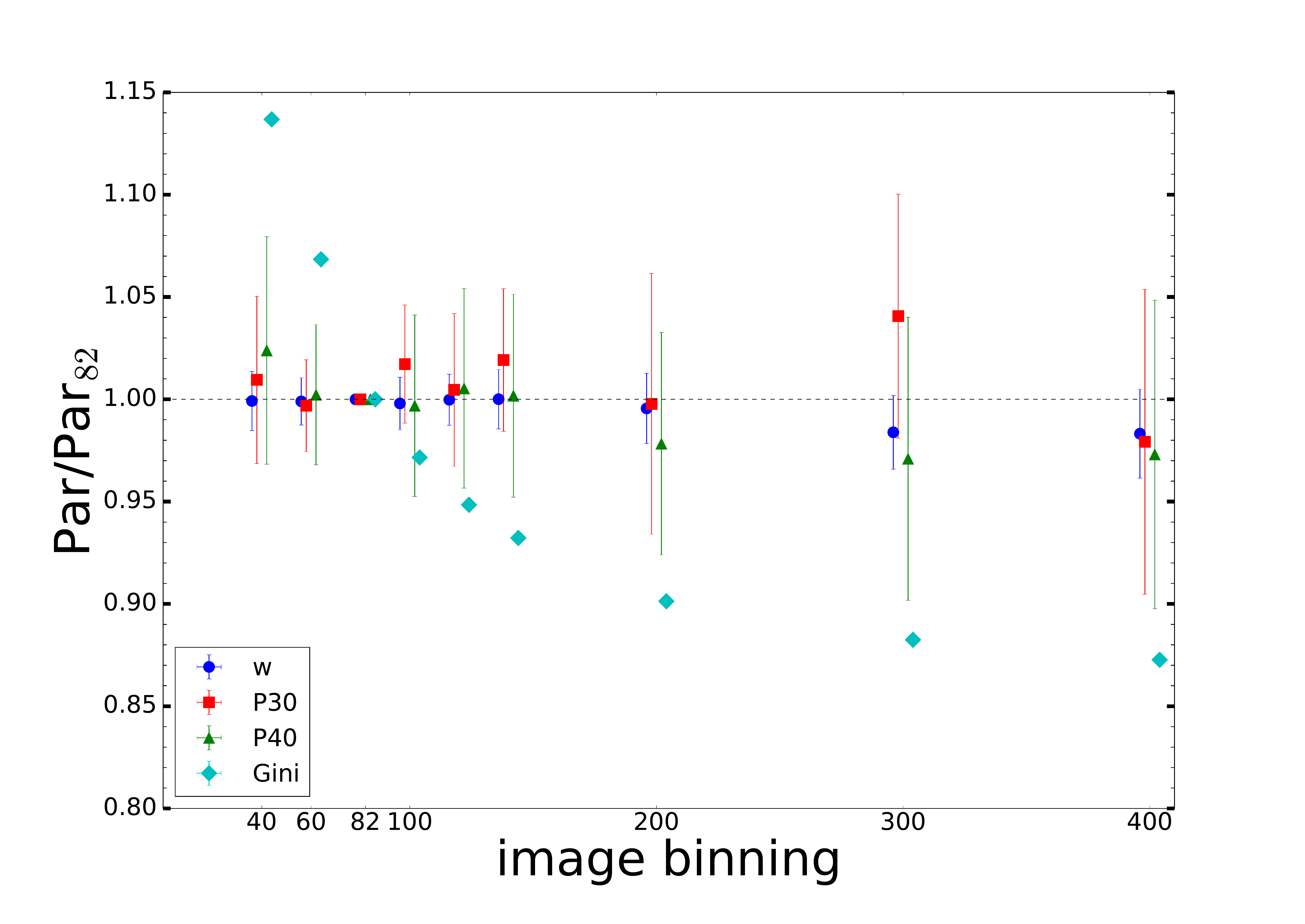}{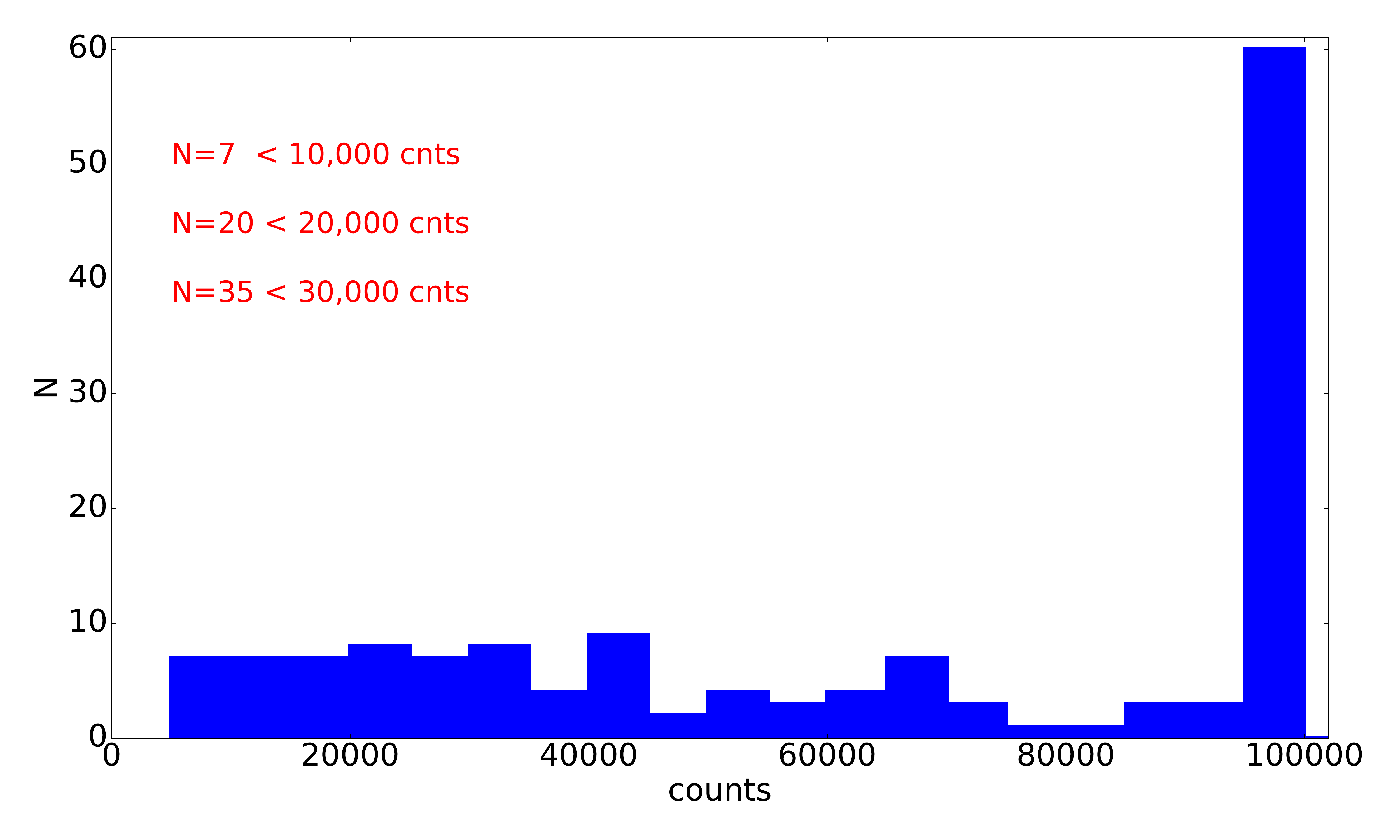}
\caption{{\it  left:} The relative change of the the centroid-shift (blue), P30 (red), P40 (green) and Gini (cyan) for different image binning with respect to the nominal binning (82 pixels corresponding to 4.1$^{\prime\prime}$). The image binning has little impact on all the morphological parameters, except the Gini coefficient. {\it  right:} Distribution of the galaxy clusters as a function of the number of counts within $R_{500}$. In the last bin we included all the objects with more than 100,000 counts. The legend report how many objects are below a certain threshold. \label{fig:ratio}}
\end{figure*}

\section{Robustness of the parameters}
The clusters in our sample cover a large redshift range so their extension on the sky varies from object to object, which might introduce systematic uncertainties in our measurements. In fact, for more distant  objects, it might be harder to measure the small scale substructures. Moreover, if the same image binning is used, for clusters at different redshift we will probe a different physical scale.  Furthermore, while some clusters have been observed with very long exposures, others have been observed with relatively short observations. This, can introduce an uncertainty in the determination of the X-ray peak (in particular for the most unrelaxed objects) because of the poorer statistic, and  reduce our ability to detect the smaller and fainter substructures. \\
In this section we describe the tests we performed to ensure that our results are stable and robust.
First, we checked how important is the choice of the image binnings on the estimated parameters. Indeed, that choice has no impact on the parameters determined using the SB profiles (e.g. central density or concentration) but may play a role for the parameters derived using the images. In Fig. A1 we show the results for 7 different binning. The centroid shift, being basically a measure of the flux distribution, is insensitive to the choice of the binning. This is very important when comparing clusters at different redshifts. The power ratios, which are a measure of the surface brightness inhomogeneities were expected to be a bit more sensitive to the choice of the image binning. In particular one would have expected to find lower values (i.e. more relaxed objects) when a higher binning is used, but the results in Fig. A1 show that this is not the case and also that the power ratios are robust and independent of the choice of binning (if that would not have been the case, we would have found more relaxed objects at high redshift due to the different physical space probed by the same binning). The Gini coefficient instead shows  a trend with the binning. In particular increasing the binning leads to a lower Gini coefficient. This happens because a larger binning tends to homogenize the flux distribution over the considered pixels.  In fact, conceptually Gini  is computed by ordering the pixels in ascending order by flux (or counts) and then comparing the resulting cumulative distribution to what would be expected if all the pixels would have the same flux. So, when the flux difference from pixel to pixel is reduced, the cumulative distribution tends to deviate less from a perfectly even flux distribution. 
As a consequence of that and given the fact that we used a constant binning for our X-ray images, the obtained Gini factors for high redshift clusters tend to be biased low with respect to the low redshift objects. 

To test if the different exposures of the clusters in our sample can systematically bias our results, we recomputed the morphological parameters for all the objects reducing the observation times by 50$\%$, 80$\%$, 90$\%$, and 95$\%$, respectively. Again, we found that the centroid shift is not sensitive to the quality of the data and the correct value can be recovered with relatively short observations. Indeed, the longer the observations, the smaller are the statistical errors associated with the measurements.  The power ratios are more sensitive to the number of source counts. For example, \cite{2013A&A...549A..19W} showed that relaxed clusters appear more disturbed if the number of counts is significantly less of 30,000 counts.  In general our clusters have a very good statistic with only a few clusters below 10,000 source counts (see the histogram in Fig. A1).  But applying a counts cut to our sample does not improve significantly the completeness and the purity of the sample. For example if we exclude all the objects with less than 30,000 counts, our completeness for P30$<$1E-7 increases from 90$\%$ to 93$\%$  and the purity from 79$\%$ to 84$\%$. \\ 
We note that the X-ray peaks determined with a different fraction of the exposure time agree within a few arcsec for all the objects and that the scatter of the parameter values due to that difference in the uncertainty of the cluster center is negligible, if compared with the errorbars. 

\begin{figure}[b!]
\figurenum{B1}
\plotone{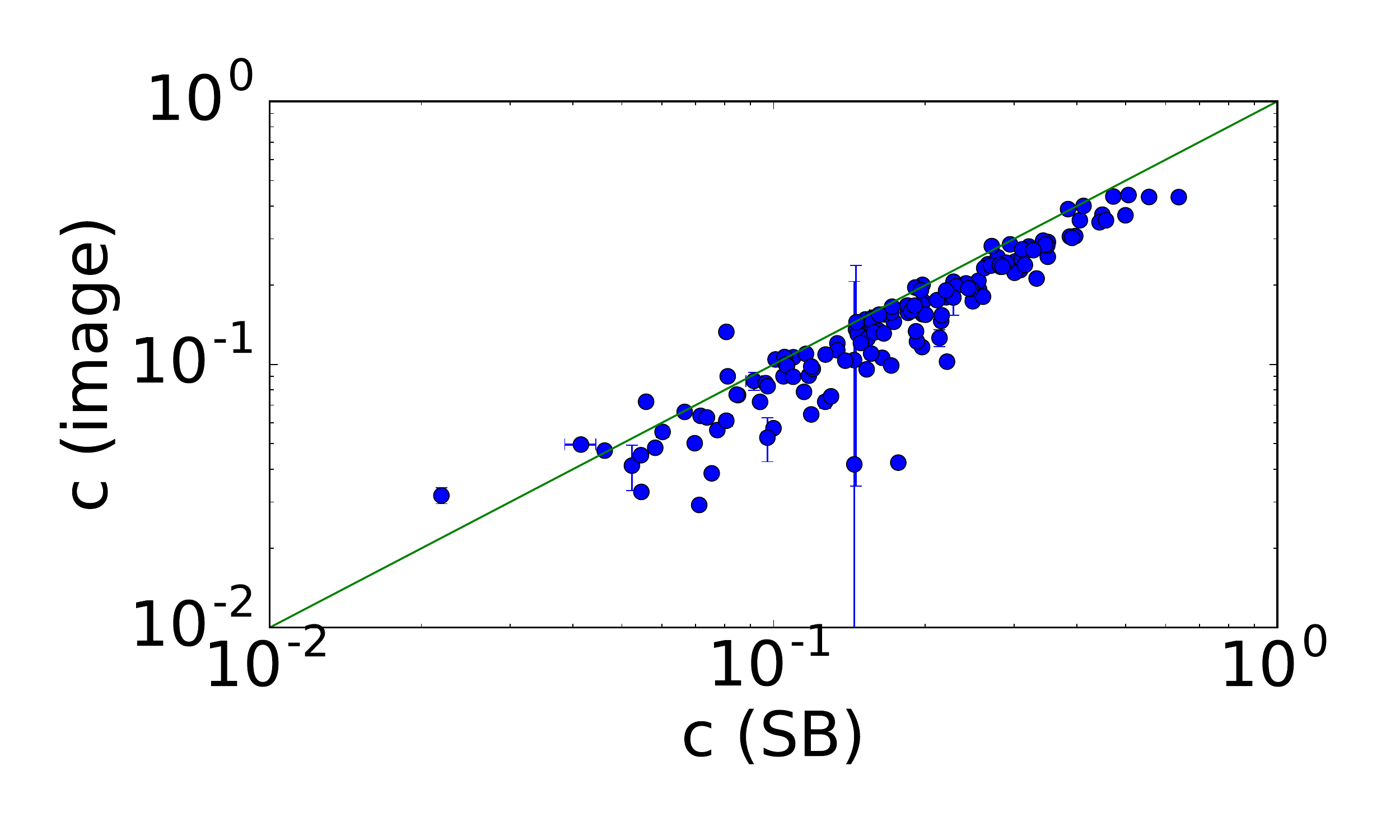}
\caption{Comparison between the concentration values computed using the images and the SB profiles.\label{fig:massmor}}
\end{figure}

\begin{table}[b!]
\renewcommand{\thetable}{\arabic{table}}
\centering
\caption{Spearman and Pearson rank test correlation and probability for no correlation    between the cluster global properties and the concentration parameter calculated using the XMM-Newton images. }
\begin{tabular}{|c|cccc|}
\hline
\hline
& \multicolumn{4}{c|}{$R_{500}$}  \\
Relation & \multicolumn1c{r} & \multicolumn1c{p-value}& \multicolumn1c{$\rho$} & \multicolumn{1}{c|}{p-value} \\
\hline
\decimals
M$_{500}$-c     & 0.03    & 0.77 & 0.01    & 0.91    \\
L$_X$-c      & 0.21    &  0.03   & 0.17   & 0.07  \\
z-c & -0.17      & 0.06    & -0.18   & 0.05  \\
\hline
\end{tabular}
\end{table}

\section{Comparison between the concentration values obtained from the images and SB profiles}
Due to the XMM-Newton PSF the surface brightness of a cluster look smoother than what it is in reality. In particular,  the effect is bigger for the more distant objects because more photons originating in the center are spread out to much larger regions. As a consequence of that using the XMM-Newton images leads to systematic lower concentrations for the more distant objects than the low redshift clusters. On the other hand, by using the SB profiles might lead to some biases for the most disturbed systems because the clear substructures have to be removed to properly fit the profiles. Depending on the region where these substructures are the effect can be either reducing or increasing the concentration because of the lower number of counts within one of the two different circular apertures.  
In Fig. B1 we show the comparison between the concentration used in the paper (i.e. c(SB)) and the concentration calculated from the images. Indeed the correlation is good and in general SB profiles give a higher concentration because the PSF is taken into account. The correlation between the cluster global properties and the concentration computed with the images give qualitatively the same results as when the SB profiles are used for the concentration.

\begin{figure*}[hp!]
\figurenum{C1}
\includegraphics[width=7in]{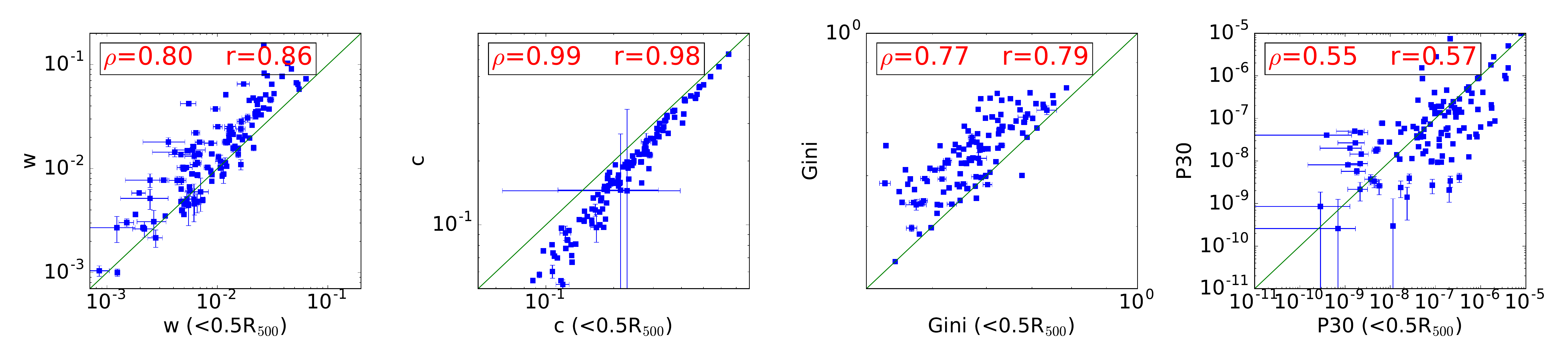}
\caption{Comparison between the morphological parameters calculated within 0.5$R_{500}$ and $R_{500}$. Note that for this comparison both {\it w} values have been renormalized by $R_{500}$ (i.e. we divided by two the value of {\it w}(R$<$R$_{500}$)). The equality line is shown in green. }
\end{figure*}

\begin{figure*}[hp!]
\figurenum{C2}
\begin{center}
\includegraphics[height=2.5in]{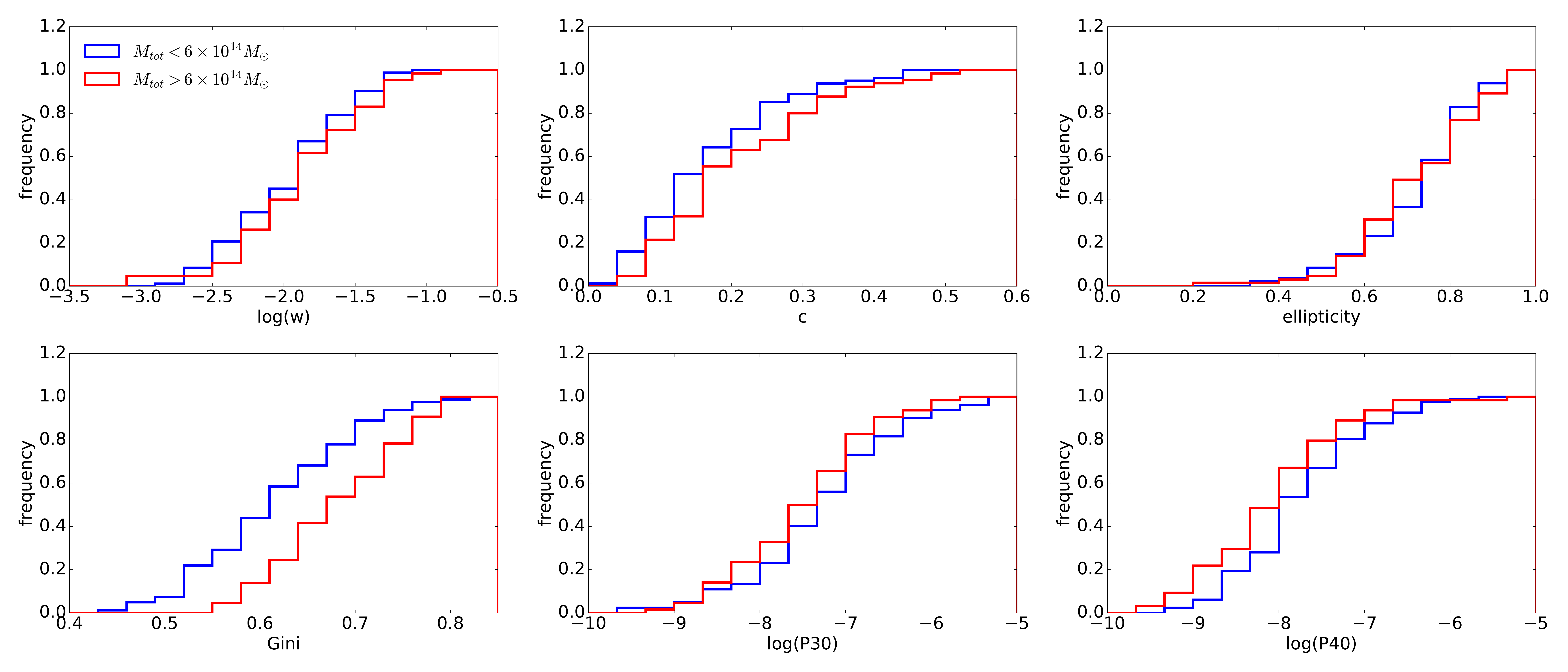}
\end{center}
\caption{Comparison of the centroid shift ({\it top-left}), concentration ({\it top-center}), ellipticity ({\it top-right}), Gini coefficient ({\it bottom-left}), P30 ({\it bottom-center}), and  P40 ({\it bottom-right}) values computed within 0.5$R_{500}$ and subdividing the sample by the total mass.}
\end{figure*}

\begin{figure*}[hp!]
\figurenum{C3}
\begin{center}
\includegraphics[height=2.5in]{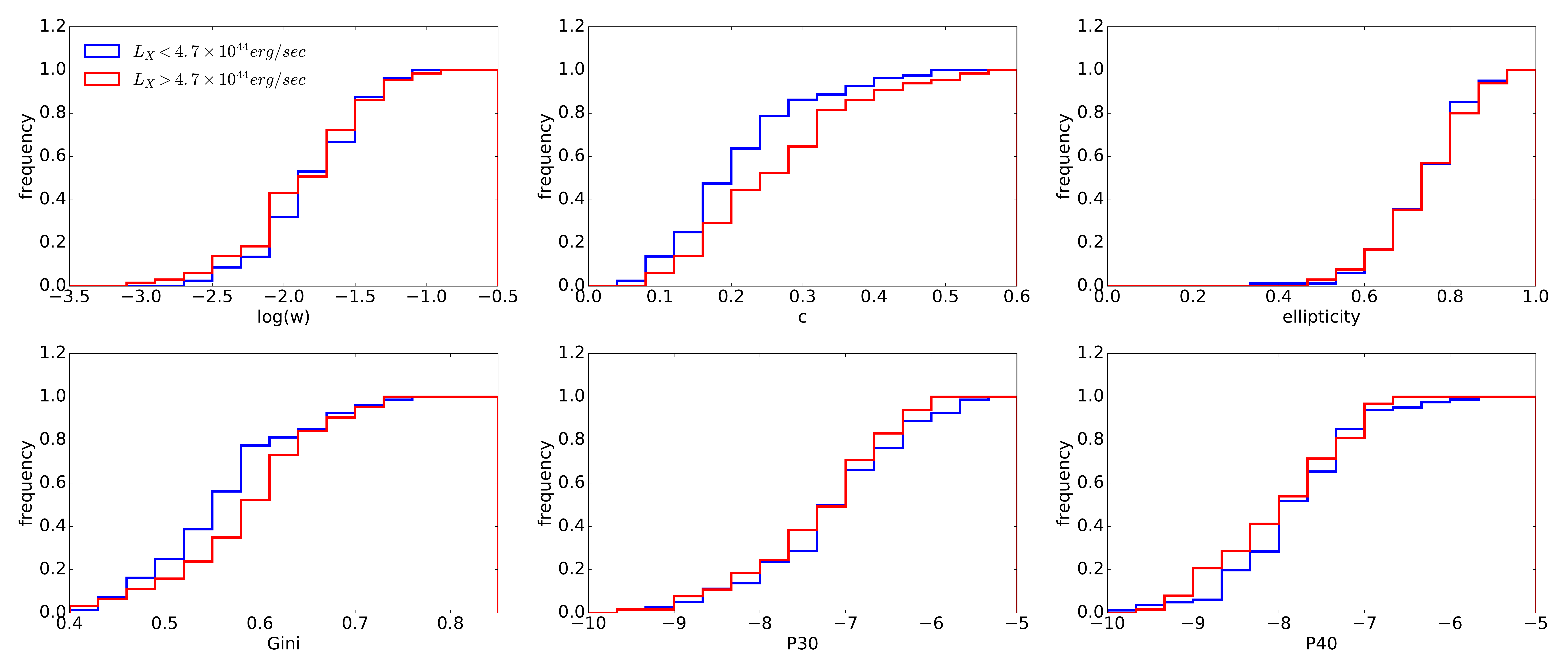}
\end{center}
\caption{Comparison of the centroid shift ({\it top-left}), concentration ({\it top-center}), ellipticity ({\it top-right}), Gini coefficient ({\it bottom-left}), P30 ({\it bottom-center}), and  P40 ({\it bottom-right}) values computed within 0.5$R_{500}$ and subdividing the sample by the total luminosity.}
\end{figure*}

\begin{figure*}[hp!]
\figurenum{C4}
\includegraphics[width=7in]{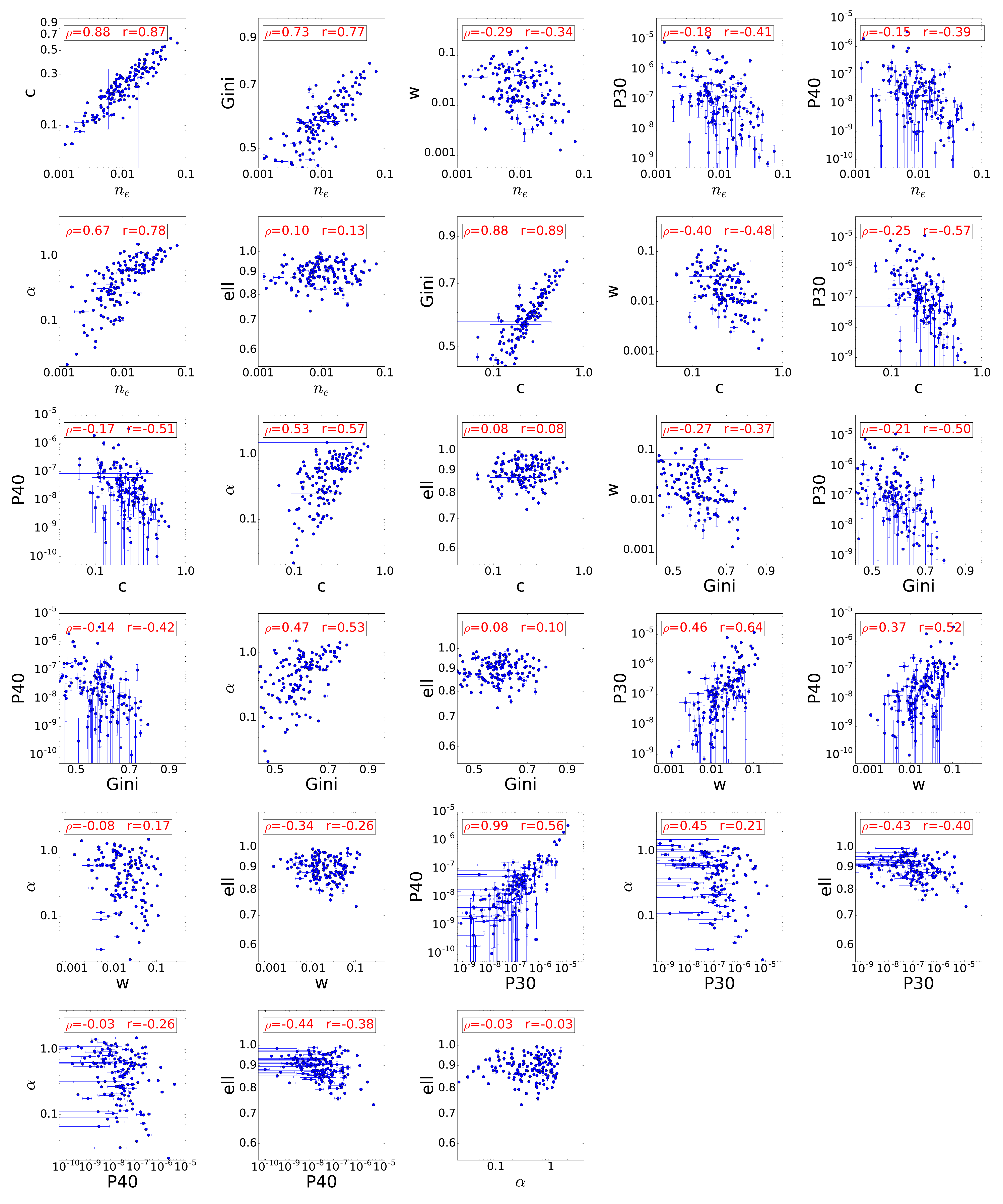}
\caption{Parameters obtained within $0.5R_{500}$ plotted in the parameter-parameter planes. $\rho$ and {\it r} values indicate the  Pearson and Spearman correlation coefficient (note that {\it r } is computed on ranks and so characterizes monotonic correlations, while $\rho$ is on true values and characterizes linear correlation). Some of the parameters show a clear and strong correlation while others are much more scattered.}
\end{figure*}

\section{Cluster properties and morphology parameters at 0.5$R_{500}$}
Clusters are continuously growing through accretion of smaller mass units. Substructures have been found to be relatively common in the outskirts of galaxy clusters (e.g. see for example the preliminary results of the XMM-Newton  Cluster Outskirt Project, \citealt{2017AN....338..293E}, \citealt{2016A&A...595A..42T}).  As a consequence, the radius at which one computes the morphological parameters may assume a relevant role. Thus, if one limits the analyses to the innermost cluster regions (e.g. within 0.5$R_{500}$), one might miss some of the infalling structures and mark a cluster as relaxed instead of disturbed. 

In  Fig. C1 we compare {\it w}, {\it c}, {\it g}, and {\it P30} calculated within 0.5R$_{500}$ and R$_{500}$ using only the clusters for which $R_{500}$ fits within the FOV. Indeed the concentration parameters show a very strong linear correlation (the Pearson rank test gives a correlation of 0.99) suggesting that the selection of clusters based on this parameter is not affected by using a different radius (i.e. the clusters with a high concentration at R$_{500}$ have also a high concentration at 0.5R$_{500}$).   This is due to the continuous and smooth shape of the SB profiles and to the fact that they are derived after the removal of all the visible substructures. The other parameters show a similar linear correlation (0.80, 0.77, and 0.55 for {\it w}, {\it g}, and {\it P30}, respectively) but are more scattered. Centroid-shift and power ratios  are more sensitive to the presence of substructures and so the choice of the radius used for the calculation has a larger impact. In fact, the centroid-shift measures the centroid variations in different aperture regions, so the presence of eventual substructures in the region 0.5-1R$_{500}$ can dramatically change the centroid position for half of the considered apertures (i.e. the 5 apertures with r$\le n \times R_{500}$, with n=6-10). Similarly for the power ratios which are based on a 2D multipole expansion of the SB distribution (representing the mass distribution) and account for the azimuthal structures.  
The multipole moments are a measure for the substructures and depend on the distance to the origin of eventual substructures, as well as their angular dependence and since they are only sensitive to structures having a scale smaller than the considered aperture (see \citealt{1995ApJ...452..522B} for more details about this last point) the choice of the radius at which P30 and P40 are calculated have impact on the results.  Considering a smaller region (e.g. 0.5$R_{500}$ instead of $R_{500}$) leads to a smaller Gini coefficient because, as explained in Appendix A,   all the low flux pixels that strongly contribute to move away the cumulative distribution from the even distribution are removed.

Despite the different parameter values at different scales the parameter-parameter relations calculated at  0.5$R_{500}$ are pretty similar to what obtained at  $R_{500}$. In Fig. C4 we show the plots as  Fig. 2,  but obtained within 0.5$R_{500}$. 

Similarly, when we compared the cluster properties with the morphological parameters computed within  0.5$R_{500}$ (see Fig. C2 and C3) we obtain qualitatively the same results as discussed in Sect. 4.4 and 5.3.

\section{Relaxed vs disturbed clusters}
In Fig. D1 we show the distribution of the clusters visually classified as relaxed, ``mix'', and disturbed for the combination of parameters that better perform in distinguishing the most relaxed and most disturbed systems (see Table 2).

\begin{figure*}[p!]
\figurenum{D1}
\includegraphics[width=7in]{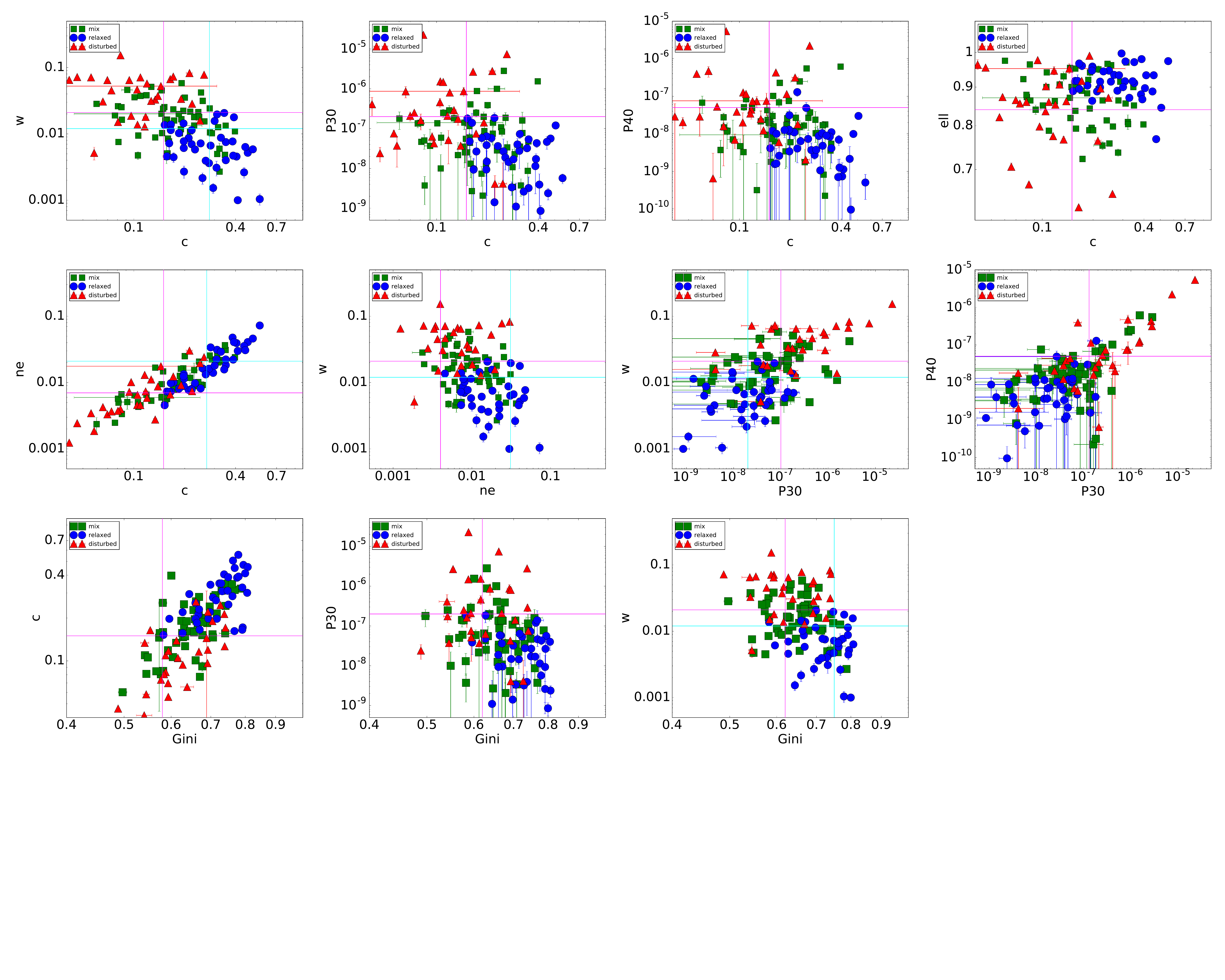}
\vspace{-98pt}
\caption{Distribution of the relaxed (blue), ``mix'' (green), and disturbed (red) clusters as a function of the different morphological parameters. The magenta and cyan lines represent the threshold values listed in Table 2 which were used to compute the completeness and the purity of the samples. }
\end{figure*}

\begin{figure*}[p!]
\figurenum{F1}
\begin{center}
\includegraphics[height=1.82in]{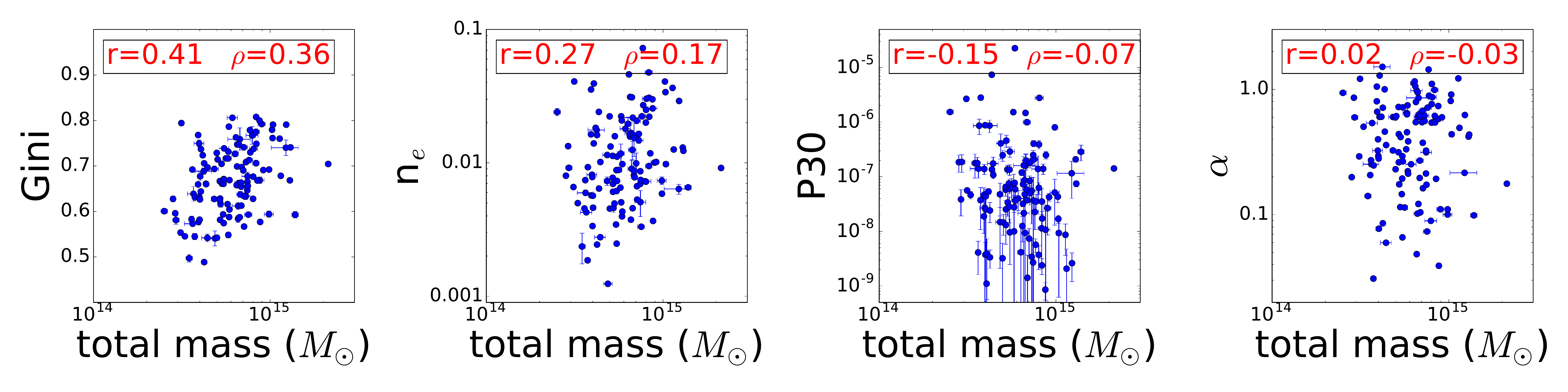}
\end{center}
\vspace{-20pt}
\caption{Correlations between the total mass and some of the morphological parameters computed within R$_{500}$. The correlation coefficient decreases from the left to the right.}
\end{figure*}

\begin{figure*}[p!]
\figurenum{F2}
\begin{center}
\includegraphics[height=1.82in]{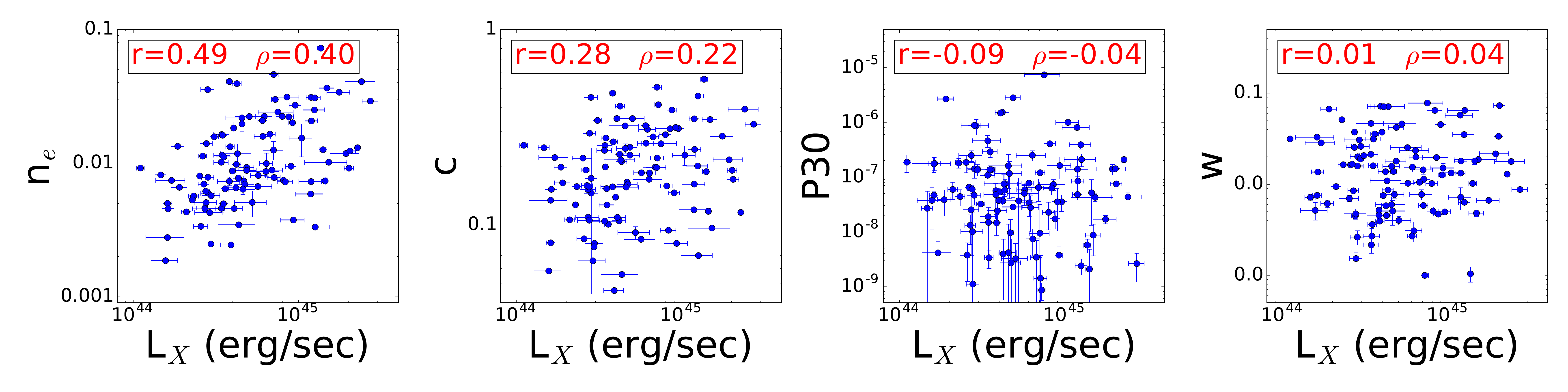}
\end{center}
\vspace{-15pt}
\caption{Correlations between the X-ray luminosity and some of the morphological parameters computed within R$_{500}$. The correlation coefficient decreases from the left to the right.}
\end{figure*}

\begin{table}[h!]
\centering
\caption{Spearman and Pearson rank test correlation and probability for no correlation    between the cluster global properties and the morphological parameters computed within 0.5$R_{500}$. } 
\begin{tabular}{|c|cccc|}
\hline
\hline
& \multicolumn{4}{c|}{$0.5R_{500}$}  \\
Relation & \multicolumn1c{r} & \multicolumn1c{p-value}& \multicolumn1c{$\rho$} & \multicolumn{1}{c|}{p-value} \\
\hline
\decimals
M$_{500}$-c     & 0.09 & 0.27 & 0.03 & 0.70  \\
M$_{500}$-w    & 0.07 & 0.38  & 0.14 & 0.09 \\
M$_{500}$-$n_e$ & 0.28   & $<$0.01 & 0.15 & 0.06 \\
M$_{500}$-Gini      & 0.10 & 0.24 & 0.05 & 0.58\\
M$_{500}$-cusp  & 0.00 & 0.97  & -0.05  & 0.59 \\
M$_{500}$-P30   & 0.10 & 0.25 & -0.07 & 0.41 \\
\hline
L$_X$-c      &  0.28 & $<0.01$ & 0.19 & 0.02  \\
L$_X$-w     &  -0.04 & 0.63 & 0.05 & 0.59  \\
L$_X$-$n_e$ & 0.50 & $<$0.01 & 0.37 & $<$0.01 \\
L$_X$-Gini   & 0.22 & $<0.01$ & 0.09 & 0.28 \\
L$_X$-cusp      & 0.14   & 0.12   & 0.11    & 0.25 \\
L$_X$-P30   & -0.05 & 0.58 & -0.09 & 0.25   \\
\hline
\end{tabular}
\end{table}

\renewcommand{\tablename}{I1}
\renewcommand{\thetable}{I1}
\begin{table}[b!]
\centering
\caption{Pearson and Spearman rank test correlation and probability for no correlation    between pairs of morphological parameters computed within $R_{500}$. } 
\begin{tabular}{|c|cccc|}
\hline
\hline
& \multicolumn{4}{c|}{$R_{500}$}  \\
Relation & \multicolumn1c{$\rho$} & \multicolumn1c{p-value}& \multicolumn1c{r} & \multicolumn{1}{c|}{p-value} \\
\hline
\decimals
$n_e-c$     & 0.87 & $<$0.01 & 0.87 & $<$0.01  \\
$n_e-Gini$ & 0.66 & $<$0.01 & 0.73 & $<$0.01  \\
$n_e-w$     & -0.29 & $<$0.01 & -0.42 & $<$0.01  \\
$n_e-P30$ & -0.06 & 0.48 & -0.34 & $<$0.01  \\
$n_e-P40$  & -0.06 & 0.50 & -0.33 & $<$0.01  \\
$n_e-\alpha$  & 0.79 & $<$0.01 & 0.81 & $<$0.01  \\
$n_e-ell$    & 0.02 & 0.85 & 0.05 & 0.60  \\
$c-Gini$ & 0.71 & $<$0.01 & 0.71 & $<$0.01  \\
$c-w$     & -0.44 & $<$0.01 & -0.53 & $<$0.01  \\
$c-P30$ & -0.11 & 0.24 & -0.39 & $<$0.01  \\
$c-P40$  & -0.10 & 0.30 & -0.37 & $<$0.01  \\
$c-\alpha$  & 0.64 & $<$0.01 & 0.60 & $<$0.01  \\
$c-ell$    & 0.08 & 0.41 & 0.08 & 0.39  \\
$Gini-w$     & -0.35 & $<$0.01 & -0.40 & $<$0.01  \\
$Gini-P30$ & -0.14 & 0.13 & -0.37 & $<$0.01  \\
$Gini-P40$  & -0.14 & 0.12 & -0.49 & $<$0.01  \\
$Gini-\alpha$  & 0.46 & $<$0.01 & 0.46 & $<$0.01  \\
$Gini-ell$    & 0.18 & 0.05 & 0.20 & 0.03  \\
$w-P30$ & 0.63 & $<$0.01 & 0.53 & $<$0.01  \\
$w-P40$  & 0.61 & $<$0.01 & 0.45 & $<$0.01  \\
$w-\alpha$  & -0.22 & 0.02 & -0.32 & $<$0.01  \\
$w-ell$    & 0.18 & 0.05 & 0.20 & 0.03  \\
$P30-P40$  & 0.99 & $<$0.01 & 0.56 & $<$0.01  \\
$P30-\alpha$  & -0.04 & 0.68 & -0.27 & $<$0.01  \\
$P30-ell$    & -0.43 & $<$0.01 & -0.40 & $<$0.01  \\
$P40-\alpha$  & -0.03 & 0.77 & -0.26 & $<$0.01  \\
$P40-ell$    & -0.44 & $<$0.01 & -0.38 & $<$0.01  \\
$\alpha-ell$   & -0.03 & 0.71 & -0.03 & 0.74  \\
\hline
\end{tabular}
\end{table}

\section{Visual classification}
The visual classification, apart from being subjective, might also depends on other criteria like, for example, the goodness of the images. We performed a few tests to ensure that our visual classification is robust. 

First of all, we checked whether the same clusters are rated similarly when showed a second time to the six astronomers. This was done by showing multiple times the same images for 20 clusters, randomly chosen and displayed in a random order. All the 20 clusters have been classified in the same way (i.e. relaxed, ``mix'', and disturbed) with an average dispersion around the mean of 0.25. In particular the most relaxed and most disturbed (i.e. the clusters with an average grade lower than 2 and greater than 3) show a smaller dispersion (0.13) than the ``mix'' objects (0.37). 

Indeed, the number of counts is also an important parameter when classifying the clusters. Some clusters have very good data quality making easier to spot eventual surface brightness features. 
Furthermore,  the image treatment can also play an important role because, for example, by progressively over-saturating the central regions of a cool core cluster may help to reveal more and more structures because the contrast in the outer regions starts to become more evident. To test whether these issues can bias the visual classification, a second image with reduced number of counts was produced and/or the color contrast changed for 40 galaxy clusters (again randomly selected).  The new images were produced to have 10 to 30 thousand of source counts (corresponding, depending on the cluster, to 10-50$\%$ of the original total number of counts).  Again, we found a very good agreement between the averaged grade obtained with the reduced and total number of counts. The dispersion around the mean was 0.14. Anyway, we note that for 62,5$\%$ of the objects the averaged grade is lower when the classification was done with reduced number of counts indicating as we said that a better data quality makes easier  to identify possible substructures. Nonetheless, only two objects were classified differently from what done with the total number of counts. Moreover, given their morphological parameter values the qualitative results of the paper would not change because they would fall in the quadrants associated with the most relaxed clusters.

\section{Correlation plots between the cluster properties and the morphological parameters}

\section{Cluster images}
Although the classification is subjective, broadly speaking, objects with circular X-ray isophotes and without substructures are classified as relaxed, objects without substructures but not perfectly circular X-ray isophotes (e.g. sloshing)  are classified as semi-relaxed,  objects with substructures but still with a well formed cluster-core  (e.g. A85)  are classified as semi-disturbed, and double or complex objects with clear evidence of merging are classified as disturbed. See the cluster images in Fig. J1.

\section{Parameter values}
All the parameter values used in this paper and calculated within R$_{500}$ are listed in Table H1. 

\section{Parameter-parameter correlations}
In Table I1 we list all the correlation coefficients and relative p-values for the plots showed in Fig. 2.

\newpage
\renewcommand{\tablename}{H1}
\renewcommand{\thetable}{H1}
\begin{table*}
\caption{Morphological parameters for the 120 galaxy clusters with $R_{500}$ completely fitting withing the XMM-Newton FOV. The last column indicates if the cluster was visually classified as relaxed (R), mix (M), or disturbed (D).} 
\resizebox{\textwidth}{!}{%
\begin{tabular}{cccccccccccc}
\hline
\hline
Planck & alternative & R$_{500}$ & $n_e$ & $cusp$ & $c$ & $w$ & $Gini$ & $P30$ & $P40$ & $ell$ & Dynamical \\
name & name & kpc & $\times10^{-2}$ & & & $\times10^{-2}$ & & $\times10^{-7}$ & $\times10^{-8}$ & & State \\	
\hline
\decimals
G000.44-41.83 & A3739 & 1114 & 0.73 $\pm$ 0.05 & 0.33 $\pm$ 0.01 & 0.16 $\pm$ 0.01 & 0.46 $\pm$ 0.06 & 0.63 $\pm$ 0.01& 0.48 $\pm$ 0.29 & 1.19 $\pm$ 0.93 & 0.92 $\pm$ 0.01 & R \\
G002.74-56.18 & RXCJ2218.6-3853 & 1106 & 1.01 $\pm$ 0.01 & 0.36 $\pm$ 0.01 & 0.24 $\pm$ 0.01 & 2.13 $\pm$ 0.04 & 0.70 $\pm$ 0.01& 1.08 $\pm$ 0.16 & 1.66 $\pm$ 0.41 & 0.81 $\pm$ 0.01 & M \\
G003.90-59.41 & A3888 & 1270 & 0.99 $\pm$ 0.20 & 0.55 $\pm$ 0.01 & 0.18 $\pm$ 0.01 & 1.99 $\pm$ 0.03 & 0.69 $\pm$ 0.01& 0.07 $\pm$ 0.04 & 1.21 $\pm$ 0.29 & 0.88 $\pm$ 0.01  & M \\
G006.70-35.54 & A3695 & 1065 & 0.57 $\pm$ 0.01 & 0.66 $\pm$ 0.01 & 0.08 $\pm$ 0.01 & 2.61 $\pm$ 0.05 & 0.58 $\pm$ 0.01& 1.39 $\pm$ 0.29 & 2.75 $\pm$ 0.80 & 0.88 $\pm$ 0.01  & M \\
G006.78+30.46 & A2163 & 1817 & 0.92 $\pm$ 0.01 & 0.18 $\pm$ 0.01 & 0.19 $\pm$ 0.01 & 3.36 $\pm$ 0.01 & 0.70 $\pm$ 0.01& 1.42 $\pm$ 0.04 & 11.40 $\pm$ 0.27 & 0.99 $\pm$ 0.01  & D \\
G008.44-56.35 & A3854 & 1061 & 1.64 $\pm$ 0.01 & 0.80 $\pm$ 0.01 & 0.25 $\pm$ 0.01 & 0.22 $\pm$ 0.04 & 0.66 $\pm$ 0.01& 0.19 $\pm$ 0.10 & 0.73 $\pm$ 0.32 & 0.92 $\pm$ 0.01  & R \\
G008.93-81.23 & A2744 & 1360 & 0.59 $\pm$ 0.01 & 0.11 $\pm$ 0.01 & 0.12 $\pm$ 0.01 & 5.74 $\pm$ 0.03 & 0.69 $\pm$ 0.01& 8.05 $\pm$ 0.21 & 7.37 $\pm$ 0.36 & 0.85 $\pm$ 0.01  & D \\
G021.09+33.25 & A2204 & 1323 & 7.25 $\pm$ 0.03 & 1.44 $\pm$ 0.01 & 0.56 $\pm$ 0.01 & 0.10 $\pm$ 0.02 & 0.78 $\pm$ 0.01& 0.06 $\pm$ 0.02 & 0.05 $\pm$ 0.03 & 0.97 $\pm$ 0.01  & R \\
G036.72+14.92 & &  1241 & 1.96 $\pm$ 0.24 & 1.12 $\pm$ 0.01 & 0.25 $\pm$ 0.01 & 1.58 $\pm$ 0.05 & 0.73 $\pm$ 0.01& 0.04 $\pm$ 0.10 & 0.20 $\pm$ 0.53 & 0.87 $\pm$ 0.01  & D \\
G039.85-39.98 & A2345 & 1077 & 0.24 $\pm$ 0.01 & 0.09 $\pm$ 0.01 & 0.05 $\pm$ 0.01 & 7.19 $\pm$ 0.05 & 0.49 $\pm$ 0.01& 0.24 $\pm$ 0.10 & 2.04 $\pm$ 0.68 & 0.95 $\pm$ 0.01  & D \\
G042.82+56.61 & A2065 & 1189 & 1.13 $\pm$ 0.01 & 0.72 $\pm$ 0.01 & 0.19 $\pm$ 0.01 & 1.66 $\pm$ 0.03 & 0.62 $\pm$ 0.01& 0.65 $\pm$ 0.09 & 1.77 $\pm$ 0.35 & 0.84 $\pm$ 0.01  & M \\
G046.08+27.18 & RXCJ1731+22 & 1148 & 0.38 $\pm$ 0.01 & 0.05 $\pm$ 0.01 & 0.08 $\pm$ 0.01 & 1.51 $\pm$ 0.18 & 0.58 $\pm$ 0.01& 1.62 $\pm$ 0.70 & 1.59 $\pm$ 1.45 & 0.86 $\pm$ 0.01  & D \\
G046.50-49.43 & A2420 & 1194 & 0.70 $\pm$ 0.01 & 0.44 $\pm$ 0.01 & 0.16 $\pm$ 0.01 & 0.85 $\pm$ 0.03 & 0.58 $\pm$ 0.01& 0.13 $\pm$ 0.07 & 1.58 $\pm$ 0.57 & 0.91 $\pm$ 0.01  & M \\
G049.20+30.86 & RXJ1720.1+2638 & 1241 & 4.61 $\pm$ 0.01 & 1.16 $\pm$ 0.01 & 0.51 $\pm$ 0.01 & 0.58 $\pm$ 0.04 & 0.76 $\pm$ 0.01& 1.20 $\pm$ 0.15 & 2.99 $\pm$ 0.61 & 0.84 $\pm$ 0.01  & R \\
G049.33+44.38 & A2175 & 1049 & 0.46 $\pm$ 0.01 & 0.21 $\pm$ 0.01 & 0.15 $\pm$ 0.01 & 1.36 $\pm$ 0.10 & 0.58 $\pm$ 0.01& 1.78 $\pm$ 0.47 & 0.41 $\pm$ 0.60 & 0.89 $\pm$ 0.01  & R \\
G049.66-49.50 & A2426 & 1090 & 1.40 $\pm$ 0.01 & 0.60 $\pm$ 0.01 & 0.29 $\pm$ 0.01 & 0.15 $\pm$ 0.03 & 0.64 $\pm$ 0.01& 0.01 $\pm$ 0.03 & 0.88 $\pm$ 0.48 & 1.00 $\pm$ 0.01  & R \\
G053.52+59.54 & A2034 & 1189 & 0.50 $\pm$ 0.01 & 0.12 $\pm$ 0.01 & 0.15 $\pm$ 0.01 & 1.60 $\pm$ 0.07 & 0.70 $\pm$ 0.01& 1.36 $\pm$ 0.24 & 0.18 $\pm$ 0.22 & 0.90 $\pm$ 0.01  & M \\
G055.60+31.86 & A2261 & 1234 & 3.10 $\pm$ 0.01 & 0.95 $\pm$ 0.01 & 0.35 $\pm$ 0.01 & 1.33 $\pm$ 0.03 & 0.75 $\pm$ 0.01& 0.48 $\pm$ 0.08 & 0.54 $\pm$ 0.16 & 0.85 $\pm$ 0.01  & M \\
G055.97-34.88 & A2355 & 1110 & 0.28 $\pm$ 0.01 & 0.06 $\pm$ 0.01 & 0.13 $\pm$ 0.01 & 3.29 $\pm$ 0.09 & 0.54 $\pm$ 0.01& 1.75 $\pm$ 0.54 & 2.49 $\pm$ 1.63 & 0.77 $\pm$ 0.01  & D \\
G056.81+36.31 & A2244 & 1098 & 1.83 $\pm$ 0.01 & 0.60 $\pm$ 0.01 & 0.35 $\pm$ 0.01 & 0.75 $\pm$ 0.02 & 0.72 $\pm$ 0.01& 0.54 $\pm$ 0.06 & 1.10 $\pm$ 0.18 & 0.91 $\pm$ 0.01  & R \\
G056.96-55.07 & &  1255 & 1.01 $\pm$ 0.01 & 0.60 $\pm$ 0.01 & 0.10 $\pm$ 0.01 & 1.87 $\pm$ 0.04 & 0.69 $\pm$ 0.01& 0.42 $\pm$ 0.07 & 3.87 $\pm$ 0.41 & 0.80 $\pm$ 0.01  & D \\
G057.26-45.35 & RXCJ2211.7-0350 & 1334 & 4.07 $\pm$ 0.03 & 0.81 $\pm$ 0.01 & 0.39 $\pm$ 0.01 & 1.78 $\pm$ 0.04 & 0.78 $\pm$ 0.01& 0.43 $\pm$ 0.10 & 0.13 $\pm$ 0.09 & 0.87 $\pm$ 0.01  & R \\
G058.28+18.59 & RXCJ1825.3+3026 & 1028 & 0.50 $\pm$ 0.01 & 0.50 $\pm$ 0.01 & 0.08 $\pm$ 0.01 & 0.76 $\pm$ 0.01 & 0.55 $\pm$ 0.01& 0.46 $\pm$ 0.07 & 1.48 $\pm$ 0.31 & 0.87 $\pm$ 0.01  & M \\
G062.42-46.41 & A2440 & 998 & 0.66 $\pm$ 0.01 & 0.29 $\pm$ 0.01 & 0.16 $\pm$ 0.01 & 6.69 $\pm$ 0.06 & 0.55 $\pm$ 0.01& 26.80 $\pm$ 1.03 & 43.00 $\pm$ 2.94 & 0.62 $\pm$ 0.01  & D \\
G067.23+67.46 & A1914 & 1334 & 2.00 $\pm$ 0.01 & 0.54 $\pm$ 0.01 & 0.32 $\pm$ 0.01 & 1.26 $\pm$ 0.01 & 0.77 $\pm$ 0.01& 0.04 $\pm$ 0.02 & 0.08 $\pm$ 0.06 & 0.97 $\pm$ 0.01  & M\\
G071.61+29.79 & &  1039 & 0.19 $\pm$ 0.01 & 0.03 $\pm$ 0.01 & 0.06 $\pm$ 0.01 & 0.52 $\pm$ 0.11 & 0.54 $\pm$ 0.01& 0.37 $\pm$ 0.27 & 2.88 $\pm$ 2.02 & 0.87 $\pm$ 0.01  & D \\
G072.63+41.46 & A2219 & 1475 & 1.26 $\pm$ 0.01 & 0.49 $\pm$ 0.01 & 0.19 $\pm$ 0.01 & 1.02 $\pm$ 0.05 & 0.68 $\pm$ 0.01& 0.02 $\pm$ 0.03 & 0.34 $\pm$ 0.22 & 0.92 $\pm$ 0.01  & M \\
G072.80-18.72 & &  1249 & 1.64 $\pm$ 0.01 & 1.06 $\pm$ 0.01 & 0.20 $\pm$ 0.01 & 2.23 $\pm$ 0.04 & 0.66 $\pm$ 0.01& 0.12 $\pm$ 0.06 & 7.53 $\pm$ 1.19 & 0.80 $\pm$ 0.01  & M \\
G073.96-27.82 & A2390 & 1492 & 3.65 $\pm$ 0.01 & 1.22 $\pm$ 0.01 & 0.35 $\pm$ 0.01 & 0.48 $\pm$ 0.04 & 0.76 $\pm$ 0.01& 0.09 $\pm$ 0.05 & 0.36 $\pm$ 0.24 & 0.90 $\pm$ 0.01  & M \\
G080.38-33.20 & A2443 & 1053 & 0.74 $\pm$ 0.01 & 0.27 $\pm$ 0.01 & 0.22 $\pm$ 0.01 & 2.85 $\pm$ 0.02 & 0.69 $\pm$ 0.01& 0.04 $\pm$ 0.02 & 1.79 $\pm$ 0.25 & 0.90 $\pm$ 0.01  & D \\
G080.99-50.90 & A2552 & 1208 & 1.54 $\pm$ 0.42 & 0.27 $\pm$ 0.01 & 0.23 $\pm$ 0.03 & 1.33 $\pm$ 0.09 & 0.65 $\pm$ 0.01& 10.01 $\pm$ 1.13 & 24.90 $\pm$ 3.74 & 0.75 $\pm$ 0.01 & M \\
G083.28-31.03 & RXCJ2228.6+2036 & 1242 & 1.18 $\pm$ 0.03 & 0.19 $\pm$ 0.01 & 0.22 $\pm$ 0.01 & 2.16 $\pm$ 0.09 & 0.67 $\pm$ 0.01& 1.40 $\pm$ 0.30 & 2.67 $\pm$ 1.01 & 0.89 $\pm$ 0.01  & M \\
G085.99+26.71 & A2302 & 1011 & 0.24 $\pm$ 0.06 & 0.14 $\pm$ 0.01 & 0.06 $\pm$ 0.01 & 2.83 $\pm$ 0.14 & 0.50 $\pm$ 0.01& 1.78 $\pm$ 0.83 & 6.76 $\pm$ 3.28 & 0.97 $\pm$ 0.01  & M \\
G086.45+15.29 & &  1270 & 2.23 $\pm$ 0.06 & 0.72 $\pm$ 0.01 & 0.29 $\pm$ 0.01 & 1.02 $\pm$ 0.06 & 0.71 $\pm$ 0.01& 0.23 $\pm$ 0.12 & 1.05 $\pm$ 0.56 & 0.92 $\pm$ 0.01  & M \\
G092.73+73.46 & A1763 & 1271 & 0.74 $\pm$ 0.01 & 0.17 $\pm$ 0.01 & 0.16 $\pm$ 0.01 & 0.51 $\pm$ 0.06 & 0.66 $\pm$ 0.01& 4.06 $\pm$ 0.52 & 10.20 $\pm$ 1.83 & 0.79 $\pm$ 0.01 & M  \\
G093.91+34.90 & A2255 & 1211 & 0.25 $\pm$ 0.01 & 0.07 $\pm$ 0.01 & 0.08 $\pm$ 0.01 & 1.90 $\pm$ 0.06 & 0.67 $\pm$ 0.01& 1.35 $\pm$ 0.30 & 0.37 $\pm$ 0.30 & 0.92 $\pm$ 0.01  & M \\
G096.87+24.21 & &  1074 & 0.12 $\pm$ 0.01 & 0.01 $\pm$ 0.01 & 0.04 $\pm$ 0.01 & 6.51 $\pm$ 0.22 & 0.54 $\pm$ 0.02& 4.10 $\pm$ 1.99 & 2.88 $\pm$ 3.42 & 0.96 $\pm$ 0.01  & D \\
G097.73+38.11 & A2218 & 1179 & 0.73 $\pm$ 0.01 & 0.27 $\pm$ 0.01 & 0.17 $\pm$ 0.01 & 1.38 $\pm$ 0.05 & 0.67 $\pm$ 0.01& 0.10 $\pm$ 0.07 & 1.00 $\pm$ 0.42 & 0.90 $\pm$ 0.01 & R  \\
G098.95+24.86 & A2312 & 995 & 0.92 $\pm$ 0.01 & 0.60 $\pm$ 0.01 & 0.26 $\pm$ 0.01 & 3.15 $\pm$ 0.09 & 0.58 $\pm$ 0.01& 1.87 $\pm$ 0.66 & 2.67 $\pm$ 1.44 & 0.96 $\pm$ 0.01  & M \\
G106.73-83.22 & A2813 & 1132 & 0.89 $\pm$ 0.01 & 0.11 $\pm$ 0.01 & 0.20 $\pm$ 0.01 & 0.77 $\pm$ 0.10 & 0.72 $\pm$ 0.01& 0.64 $\pm$ 0.28 & 3.12 $\pm$ 1.19 & 0.95 $\pm$ 0.01  & R \\
G107.11+65.31 & A1758 & 1186 & 0.65 $\pm$ 0.01 & 0.33 $\pm$ 0.01 & 0.11 $\pm$ 0.01 & 1.37 $\pm$ 0.05 & 0.62 $\pm$ 0.01& 15.30 $\pm$ 0.87 & 12.60 $\pm$ 1.56 & 0.90 $\pm$ 0.01  & D \\
G113.82+44.35 & A1895 & 1139 & 0.46 $\pm$ 0.03 & 0.17 $\pm$ 0.01 & 0.10 $\pm$ 0.01 & 4.65 $\pm$ 0.14 & 0.62 $\pm$ 0.01& 4.60 $\pm$ 1.15 & 2.00 $\pm$ 1.63 & 0.84 $\pm$ 0.01  & D \\
G124.21-36.48 & A115N & 1072 & 2.41 $\pm$ 0.01 & 1.00 $\pm$ 0.01 & 0.26 $\pm$ 0.01 & 7.80 $\pm$ 0.04 & 0.66 $\pm$ 0.01& 74.10 $\pm$ 1.40 & 222.00 $\pm$ 5.14 & 0.65 $\pm$ 0.01  & D \\
G125.70+53.85 & A1576 & 1197 & 1.25 $\pm$ 0.20 & 0.60 $\pm$ 0.01 & 0.20 $\pm$ 0.01 & 1.44 $\pm$ 0.15 & 0.66 $\pm$ 0.01& 0.09 $\pm$ 0.22 & 1.29 $\pm$ 1.81 & 0.95 $\pm$ 0.01  & R \\
G139.19+56.35 & A1351 & 1228 & 0.51 $\pm$ 0.11 & 0.31 $\pm$ 0.01 & 0.09 $\pm$ 0.01 & 4.60 $\pm$ 0.20 & 0.68 $\pm$ 0.01& 0.37 $\pm$ 0.60 & 0.71 $\pm$ 1.60 & 0.84 $\pm$ 0.01  & M \\
G149.73+34.69 & A0665 & 1353 & 0.95 $\pm$ 0.01 & 0.56 $\pm$ 0.01 & 0.15 $\pm$ 0.01 & 4.52 $\pm$ 0.09 & 0.71 $\pm$ 0.01& 0.35 $\pm$ 0.25 & 4.27 $\pm$ 1.76 & 0.95 $\pm$ 0.01  & M \\
G157.43+30.33 & &  1125 & 1.68 $\pm$ 0.01 & 0.98 $\pm$ 0.01 & 0.13 $\pm$ 0.01 & 0.88 $\pm$ 0.07 & 0.61 $\pm$ 0.01& 0.22 $\pm$ 0.23 & 2.14 $\pm$ 1.83 & 0.93 $\pm$ 0.01  & M \\
G159.85-73.47 & A0209 & 1245 & 0.81 $\pm$ 0.04 & 0.20 $\pm$ 0.01 & 0.18 $\pm$ 0.01 & 1.02 $\pm$ 0.07 & 0.66 $\pm$ 0.01& 0.58 $\pm$ 0.15 & 1.26 $\pm$ 0.47 & 0.90 $\pm$ 0.01 & R  \\
G164.18-38.89 & A0399 & 1119 & 0.64 $\pm$ 0.01 & 0.46 $\pm$ 0.01 & 0.10 $\pm$ 0.01 & 3.56 $\pm$ 0.04 & 0.66 $\pm$ 0.01& 1.36 $\pm$ 0.18 & 0.47 $\pm$ 0.27 & 0.85 $\pm$ 0.01  & M \\
G166.13+43.39 & A0773 & 1250 & 0.78 $\pm$ 0.01 & 0.10 $\pm$ 0.01 & 0.22 $\pm$ 0.01 & 1.14 $\pm$ 0.06 & 0.70 $\pm$ 0.01& 0.01 $\pm$ 0.02 & 0.41 $\pm$ 0.24 & 0.92 $\pm$ 0.01  & R \\
G167.65+17.64 & &  1299 & 0.69 $\pm$ 0.01 & 0.21 $\pm$ 0.01 & 0.16 $\pm$ 0.01 & 0.77 $\pm$ 0.07 & 0.65 $\pm$ 0.01& 0.03 $\pm$ 0.05 & 0.63 $\pm$ 0.40 & 0.90 $\pm$ 0.01  & M \\
G171.94-40.65 & &  1408 & 0.98 $\pm$ 0.01 & 0.44 $\pm$ 0.01 & 0.17 $\pm$ 0.01 & 1.14 $\pm$ 0.07 & 0.79 $\pm$ 0.01& 0.09 $\pm$ 0.09 & 0.16 $\pm$ 0.25 & 0.97 $\pm$ 0.01  & R \\
G180.24+21.04 & &  1358 & 1.30 $\pm$ 0.04 & 0.42 $\pm$ 0.01 & 0.12 $\pm$ 0.01 & 1.29 $\pm$ 0.05 & 0.67 $\pm$ 0.01& 2.09 $\pm$ 0.24 & 3.46 $\pm$ 0.60 & 0.78 $\pm$ 0.01  & D \\
G182.44-28.29 & A0478 & 1415 & 3.00 $\pm$ 0.02 & 0.74 $\pm$ 0.01 & 0.41 $\pm$ 0.01 & 0.10 $\pm$ 0.01 & 0.80 $\pm$ 0.01& 0.01 $\pm$ 0.01 & 0.11 $\pm$ 0.02 & 0.93 $\pm$ 0.01  & R \\
G182.63+55.82 & A0963 & 1126 & 2.23 $\pm$ 0.01 & 0.60 $\pm$ 0.01 & 0.35 $\pm$ 0.01 & 0.40 $\pm$ 0.04 & 0.73 $\pm$ 0.01& 0.03 $\pm$ 0.03 & 0.41 $\pm$ 0.24 & 0.97 $\pm$ 0.01  & R \\
G186.39+37.25 & A0697 & 1280 & 0.73 $\pm$ 0.01 & 0.09 $\pm$ 0.01 & 0.16 $\pm$ 0.01 & 0.72 $\pm$ 0.20 & 0.77 $\pm$ 0.02& 1.39 $\pm$ 1.01 & 0.16 $\pm$ 1.07 & 0.92 $\pm$ 0.02  & R \\
G195.62+44.05 & A0781 & 1105 & 0.34 $\pm$ 0.01 & 0.15 $\pm$ 0.01 & 0.06 $\pm$ 0.01 & 7.11 $\pm$ 0.05 & 0.59 $\pm$ 0.01& 0.75 $\pm$ 0.14 & 39.50 $\pm$ 2.17 & 0.82 $\pm$ 0.01  & D \\
G195.77-24.30 & A0520 & 1314 & 0.72 $\pm$ 0.01 & 0.76 $\pm$ 0.01 & 0.09 $\pm$ 0.01 & 6.46 $\pm$ 0.05 & 0.63 $\pm$ 0.01& 0.63 $\pm$ 0.15 & 0.69 $\pm$ 0.29 & 0.98 $\pm$ 0.01  & D \\
G218.85+35.50 & A0750 & 1050 & 0.93 $\pm$ 0.01 & 0.25 $\pm$ 0.01 & 0.25 $\pm$ 0.01 & 4.21 $\pm$ 0.09 & 0.63 $\pm$ 0.01& 28.20 $\pm$ 1.94 & 54.90 $\pm$ 4.61 & 0.76 $\pm$ 0.01  & M \\
G225.92-19.99 & &  1187 & 3.03 $\pm$ 0.02 & 0.87 $\pm$ 0.01 & 0.21 $\pm$ 0.01 & 8.25 $\pm$ 0.12 & 0.74 $\pm$ 0.01& 27.90 $\pm$ 1.65 & 31.50 $\pm$ 4.45 & 0.76 $\pm$ 0.01  & D \\
\hline
\end{tabular}
}
\end{table*}

\renewcommand{\tablename}{H1}
\renewcommand{\thetable}{H1}
\begin{table*}
\caption{continued from previous page} 
\resizebox{\textwidth}{!}{%
\begin{tabular}{cccccccccccc}
\hline
\hline
Planck & alternative & R$_{500}$ & $n_e$ & $cusp$ & $c$ & $w$ & $Gini$ & $P30$ & $P40$ & $ell$ & Dynamical \\
name & name & kpc & $\times10^{-2}$ & & & $\times10^{-2}$ & & $\times10^{-7}$ & $\times10^{-8}$ & & State \\	
\hline
\decimals
G226.17-21.91 & A0550 & 1087 & 0.57 $\pm$ 0.02 & 0.28 $\pm$ 0.01 & 0.16 $\pm$ 0.01 & 1.64 $\pm$ 0.04 & 0.63 $\pm$ 0.01& 0.44 $\pm$ 0.14 & 2.99 $\pm$ 0.88 & 0.84 $\pm$ 0.01  & M \\
G226.24+76.76 & A1413 & 1215 & 2.08 $\pm$ 0.01 & 0.74 $\pm$ 0.01 & 0.32 $\pm$ 0.01 & 0.27 $\pm$ 0.01 & 0.79 $\pm$ 0.01& 0.77 $\pm$ 0.02 & 1.77 $\pm$ 0.07 & 0.80 $\pm$ 0.01  & M \\
G228.15+75.19 & &  1239 & 0.73 $\pm$ 0.05 & 0.10 $\pm$ 0.01 & 0.12 $\pm$ 0.01 & 1.80 $\pm$ 0.15 & 0.59 $\pm$ 0.01& 0.52 $\pm$ 0.39 & 4.41 $\pm$ 1.97 & 0.95 $\pm$ 0.01  & D \\
G228.49+53.12 & &  1061 & 3.55 $\pm$ 0.01 & 1.05 $\pm$ 0.01 & 0.45 $\pm$ 0.01 & 0.26 $\pm$ 0.05 & 0.77 $\pm$ 0.01& 0.46 $\pm$ 0.16 & 0.21 $\pm$ 0.24 & 0.89 $\pm$ 0.01  & R \\
G229.21-17.24 & &  1136 & 0.45 $\pm$ 0.01 & 0.23 $\pm$ 0.01 & 0.11 $\pm$ 0.01 & 3.74 $\pm$ 0.10 & 0.54 $\pm$ 0.01& 2.48 $\pm$ 0.56 & 8.34 $\pm$ 2.53 & 0.79 $\pm$ 0.01  & M \\
G229.94+15.29 & A0644 & 1289 & 1.57 $\pm$ 0.01 & 0.55 $\pm$ 0.01 & 0.34 $\pm$ 0.01 & 2.06 $\pm$ 0.01 & 0.70 $\pm$ 0.01& 0.32 $\pm$ 0.04 & 0.83 $\pm$ 0.13 & 0.92 $\pm$ 0.01  & R \\
G236.95-26.67 & A3364 & 1206 & 0.77 $\pm$ 0.01 & 0.31 $\pm$ 0.01 & 0.21 $\pm$ 0.01 & 0.87 $\pm$ 0.04 & 0.67 $\pm$ 0.01& 0.58 $\pm$ 0.14 & 1.09 $\pm$ 0.36 & 0.89 $\pm$ 0.01  & R \\
G241.74-30.88 & RXCJ0532.9-3701 & 1159 & 2.23 $\pm$ 0.02 & 0.63 $\pm$ 0.01 & 0.31 $\pm$ 0.01 & 0.31 $\pm$ 0.08 & 0.73 $\pm$ 0.01& 0.28 $\pm$ 0.18 & 1.04 $\pm$ 0.71 & 0.92 $\pm$ 0.01  & R \\
G241.77-24.00 & A3378 & 1070 & 3.94 $\pm$ 0.01 & 1.29 $\pm$ 0.01 & 0.41 $\pm$ 0.01 & 0.46 $\pm$ 0.02 & 0.74 $\pm$ 0.01& 0.04 $\pm$ 0.03 & 0.07 $\pm$ 0.10 & 0.90 $\pm$ 0.01  & R \\
G241.97+14.85 & A3411 & 1254 & 0.46 $\pm$ 0.01 & 0.12 $\pm$ 0.01 & 0.11 $\pm$ 0.01 & 7.11 $\pm$ 0.03 & 0.59 $\pm$ 0.01& 14.80 $\pm$ 0.37 & 11.50 $\pm$ 0.74 & 0.86 $\pm$ 0.01  & D \\
G244.34-32.13 & RXCJ0528.9-3927 & 1229 & 2.07 $\pm$ 0.06 & 0.67 $\pm$ 0.01 & 0.24 $\pm$ 0.01 & 1.80 $\pm$ 0.08 & 0.74 $\pm$ 0.01& 0.84 $\pm$ 0.19 & 0.18 $\pm$ 0.21 & 0.97 $\pm$ 0.01 & M  \\
G244.69+32.49 & A0868 & 1069 & 0.46 $\pm$ 0.01 & 0.08 $\pm$ 0.01 & 0.15 $\pm$ 0.01 & 2.54 $\pm$ 0.15 & 0.68 $\pm$ 0.01& 0.25 $\pm$ 0.21 & 2.21 $\pm$ 1.42 & 0.90 $\pm$ 0.01  & M \\
G247.17-23.32 & ABELLS0579 & 1031 & 0.60 $\pm$ 0.03 & 0.25 $\pm$ 0.01 & 0.15 $\pm$ 0.10 & 2.01 $\pm$ 0.09 & 0.57 $\pm$ 0.01& 1.41 $\pm$ 0.47 & 0.94 $\pm$ 0.96 & 0.87 $\pm$ 0.01  & M \\
G249.87-39.86 & A3292 & 948 & 0.80 $\pm$ 0.01 & 0.20 $\pm$ 0.01 & 0.22 $\pm$ 0.01 & 0.70 $\pm$ 0.05 & 0.63 $\pm$ 0.01& 1.85 $\pm$ 0.65 & 12.90 $\pm$ 3.02 & 0.90 $\pm$ 0.01  & R \\
G250.90-36.25 & A3322 & 1155 & 0.88 $\pm$ 0.01 & 0.19 $\pm$ 0.01 & 0.23 $\pm$ 0.01 & 1.79 $\pm$ 0.06 & 0.66 $\pm$ 0.01& 0.28 $\pm$ 0.14 & 1.50 $\pm$ 0.73 & 0.86 $\pm$ 0.01  & M \\
G252.96-56.05 & A3112 & 1006 & 4.08 $\pm$ 0.01 & 1.21 $\pm$ 0.01 & 0.47 $\pm$ 0.01 & 0.52 $\pm$ 0.01 & 0.79 $\pm$ 0.01& 0.56 $\pm$ 0.03 & 0.99 $\pm$ 0.07 & 0.77 $\pm$ 0.01  & R \\
G253.47-33.72 & A3343 & 1118 & 1.15 $\pm$ 0.01 & 0.60 $\pm$ 0.01 & 0.20 $\pm$ 0.01 & 0.27 $\pm$ 0.06 & 0.69 $\pm$ 0.01& 0.15 $\pm$ 0.13 & 1.16 $\pm$ 0.84 & 0.96 $\pm$ 0.01  & R \\
G256.45-65.71 & A3017 & 1143 & 1.58 $\pm$ 0.02 & 0.42 $\pm$ 0.01 & 0.26 $\pm$ 0.01 & 1.54 $\pm$ 0.05 & 0.66 $\pm$ 0.01& 0.34 $\pm$ 0.15 & 3.69 $\pm$ 0.92 & 0.80 $\pm$ 0.01  & M \\
G257.34-22.18 & A3399 & 1168 & 1.11 $\pm$ 0.02 & 0.72 $\pm$ 0.01 & 0.13 $\pm$ 0.01 & 3.14 $\pm$ 0.14 & 0.74 $\pm$ 0.01& 2.90 $\pm$ 0.61 & 7.36 $\pm$ 2.45 & 0.91 $\pm$ 0.01  & D \\
G260.03-63.44 & RXCJ0232.2-4420 & 1196 & 3.12 $\pm$ 0.01 & 0.79 $\pm$ 0.01 & 0.31 $\pm$ 0.01 & 1.97 $\pm$ 0.04 & 0.75 $\pm$ 0.01& 0.73 $\pm$ 0.17 & 0.48 $\pm$ 0.33 & 0.97 $\pm$ 0.01  & R \\
G262.25-35.36 & RXCJ0516.7-5430 & 1247 & 0.33 $\pm$ 0.01 & 0.07 $\pm$ 0.01 & 0.07 $\pm$ 0.01 & 6.47 $\pm$ 0.13 & 0.59 $\pm$ 0.01& 2.09 $\pm$ 0.54 & 0.07 $\pm$ 0.24 & 0.86 $\pm$ 0.01  & D \\
G262.71-40.91 & &  1123 & 1.80 $\pm$ 0.02 & 0.25 $\pm$ 0.01 & 0.30 $\pm$ 0.01 & 1.56 $\pm$ 0.06 & 0.81 $\pm$ 0.01& 0.40 $\pm$ 0.16 & 0.11 $\pm$ 0.15 & 0.90 $\pm$ 0.01  & R \\
G263.16-23.41 & AbellS0592 & 1294 & 2.71 $\pm$ 0.01 & 0.89 $\pm$ 0.01 & 0.31 $\pm$ 0.01 & 0.50 $\pm$ 0.03 & 0.73 $\pm$ 0.01& 0.35 $\pm$ 0.12 & 1.69 $\pm$ 0.44 & 0.86 $\pm$ 0.01  & M \\
G263.66-22.53 & A3404 & 1297 & 1.65 $\pm$ 0.02 & 0.56 $\pm$ 0.01 & 0.28 $\pm$ 0.01 & 1.05 $\pm$ 0.04 & 0.71 $\pm$ 0.01& 0.03 $\pm$ 0.03 & 1.03 $\pm$ 0.44 & 0.86 $\pm$ 0.01  & M \\
G266.03-21.25 & &  1499 & 1.23 $\pm$ 0.01 & 0.44 $\pm$ 0.01 & 0.17 $\pm$ 0.01 & 7.30 $\pm$ 0.03 & 0.74 $\pm$ 0.01& 0.75 $\pm$ 0.09 & 0.60 $\pm$ 0.17 & 0.92 $\pm$ 0.01  & D \\
G269.31-49.87 & A3126 & 1098 & 0.81 $\pm$ 0.01 & 0.32 $\pm$ 0.01 & 0.25 $\pm$ 0.01 & 0.72 $\pm$ 0.05 & 0.75 $\pm$ 0.01& 0.27 $\pm$ 0.28 & 4.89 $\pm$ 1.97 & 0.95 $\pm$ 0.01 & R  \\
G271.19-30.96 & &  1250 & 4.76 $\pm$ 0.08 & 0.99 $\pm$ 0.01 & 0.38 $\pm$ 0.01 & 0.77 $\pm$ 0.05 & 0.78 $\pm$ 0.01& 0.11 $\pm$ 0.09 & 0.07 $\pm$ 0.14 & 0.88 $\pm$ 0.01  & R \\
G271.50-56.55 & S0295 & 1200 & 0.90 $\pm$ 0.14 & 0.10 $\pm$ 0.01 & 0.19 $\pm$ 0.01 & 5.80 $\pm$ 0.09 & 0.66 $\pm$ 0.01& 0.86 $\pm$ 0.46 & 2.60 $\pm$ 1.31 & 0.94 $\pm$ 0.01  & M \\
G272.10-40.15 & A3266 & 1316 & 0.87 $\pm$ 0.01 & 0.61 $\pm$ 0.01 & 0.14 $\pm$ 0.01 & 3.73 $\pm$ 0.01 & 0.61 $\pm$ 0.01& 0.37 $\pm$ 0.03 & 1.23 $\pm$ 0.12 & 0.86 $\pm$ 0.01  & D \\
G277.75-51.73 & &  1238 & 0.37 $\pm$ 0.01 & 0.04 $\pm$ 0.01 & 0.07 $\pm$ 0.01 & 4.52 $\pm$ 0.07 & 0.58 $\pm$ 0.01& 2.50 $\pm$ 0.38 & 5.28 $\pm$ 0.96 & 0.86 $\pm$ 0.01  & D \\
G278.60+39.17 & A1300 & 1268 & 2.50 $\pm$ 0.01 & 1.10 $\pm$ 0.01 & 0.20 $\pm$ 0.01 & 3.51 $\pm$ 0.08 & 0.68 $\pm$ 0.01& 3.90 $\pm$ 0.68 & 0.60 $\pm$ 0.69 & 0.79 $\pm$ 0.01  & M \\
G280.19+47.81 & A1391 & 1201 & 0.51 $\pm$ 0.01 & 0.45 $\pm$ 0.01 & 0.11 $\pm$ 0.01 & 0.47 $\pm$ 0.06 & 0.55 $\pm$ 0.01& 0.10 $\pm$ 0.12 & 0.32 $\pm$ 0.59 & 0.90 $\pm$ 0.01  & M \\
G282.49+65.17 & ZwCl1215 & 1212 & 0.61 $\pm$ 0.02 & 0.29 $\pm$ 0.01 & 0.16 $\pm$ 0.01 & 0.45 $\pm$ 0.02 & 0.57 $\pm$ 0.01& 0.60 $\pm$ 0.10 & 1.05 $\pm$ 0.26 & 0.95 $\pm$ 0.01  & M \\
G283.16-22.93 & &  1137 & 1.62 $\pm$ 0.01 & 0.63 $\pm$ 0.01 & 0.19 $\pm$ 0.01 & 1.18 $\pm$ 0.11 & 0.64 $\pm$ 0.01& 1.98 $\pm$ 0.65 & 1.73 $\pm$ 1.60 & 0.79 $\pm$ 0.01 & M  \\
G284.46+52.43 & RXJ1206.2-0848 & 1308 & 3.39 $\pm$ 0.01 & 0.91 $\pm$ 0.01 & 0.28 $\pm$ 0.01 & 0.67 $\pm$ 0.02 & 0.76 $\pm$ 0.01& 0.17 $\pm$ 0.03 & 0.37 $\pm$ 0.09 & 0.93 $\pm$ 0.01  & R \\
G284.99-23.70 & &  1266 & 2.56 $\pm$ 0.02 & 0.58 $\pm$ 0.01 & 0.28 $\pm$ 0.01 & 2.42 $\pm$ 0.11 & 0.67 $\pm$ 0.01& 0.11 $\pm$ 0.20 & 2.37 $\pm$ 1.53 & 0.74 $\pm$ 0.01  & M \\
G285.63-17.24 & &  1007 & 1.76 $\pm$ 0.10 & 1.51 $\pm$ 0.01 & 0.14 $\pm$ 0.17 & 5.25 $\pm$ 0.16 & 0.69 $\pm$ 0.01& 8.66 $\pm$ 2.24 & 7.55 $\pm$ 4.53 & 0.95 $\pm$ 0.01  & D \\
G286.58-31.25 & &  1141 & 0.61 $\pm$ 0.01 & 0.26 $\pm$ 0.01 & 0.15 $\pm$ 0.01 & 1.03 $\pm$ 0.06 & 0.63 $\pm$ 0.01& 0.25 $\pm$ 0.12 & 2.33 $\pm$ 0.79 & 0.82 $\pm$ 0.01  & M \\
G286.99+32.91 & &  1476 & 0.66 $\pm$ 0.01 & 0.10 $\pm$ 0.01 & 0.12 $\pm$ 0.01 & 3.82 $\pm$ 0.11 & 0.59 $\pm$ 0.01& 2.87 $\pm$ 0.92 & 4.82 $\pm$ 1.78 & 0.86 $\pm$ 0.01  & M \\
G288.61-37.65 & A3186 & 1301 & 0.53 $\pm$ 0.01 & 0.16 $\pm$ 0.01 & 0.13 $\pm$ 0.01 & 5.12 $\pm$ 0.07 & 0.63 $\pm$ 0.01& 1.81 $\pm$ 0.37 & 0.03 $\pm$ 0.13 & 0.91 $\pm$ 0.01  & M \\
G292.51+21.98 & &  1147 & 0.40 $\pm$ 0.01 & 0.22 $\pm$ 0.01 & 0.08 $\pm$ 0.01 & 15.31 $\pm$ 0.11 & 0.59 $\pm$ 0.01& 227.00 $\pm$ 5.21 & 542.00 $\pm$ 19.30 & 0.67 $\pm$ 0.01  & D \\
G294.66-37.02 & RXCJ0303.8-7752 & 1253 & 0.86 $\pm$ 0.01 & 0.32 $\pm$ 0.01 & 0.17 $\pm$ 0.01 & 2.34 $\pm$ 0.09 & 0.67 $\pm$ 0.01& 2.50 $\pm$ 0.54 & 6.84 $\pm$ 1.86 & 0.89 $\pm$ 0.01  & M \\
G304.67-31.66 & A4023 & 1020 & 0.43 $\pm$ 0.01 & 0.54 $\pm$ 0.01 & 0.07 $\pm$ 0.01 & 3.07 $\pm$ 0.22 & 0.64 $\pm$ 0.02& 8.63 $\pm$ 2.72 & 47.10 $\pm$ 14.80 & 0.71 $\pm$ 0.01  & D \\
G304.84-41.42 & &  1184 & 1.47 $\pm$ 0.07 & 0.62 $\pm$ 0.01 & 0.15 $\pm$ 0.01 & 2.40 $\pm$ 0.15 & 0.63 $\pm$ 0.01& 0.56 $\pm$ 0.42 & 1.97 $\pm$ 1.73 & 0.90 $\pm$ 0.01  & M \\
G306.68+61.06 & A1650 & 1102 & 1.62 $\pm$ 0.01 & 0.71 $\pm$ 0.01 & 0.28 $\pm$ 0.01 & 0.36 $\pm$ 0.01 & 0.71 $\pm$ 0.01& 0.03 $\pm$ 0.01 & 0.27 $\pm$ 0.06 & 0.89 $\pm$ 0.01  & R \\
G306.80+58.60 & A1651 & 1181 & 1.33 $\pm$ 0.01 & 0.61 $\pm$ 0.01 & 0.27 $\pm$ 0.01 & 0.39 $\pm$ 0.03 & 0.71 $\pm$ 0.01& 0.14 $\pm$ 0.06 & 0.37 $\pm$ 0.20 & 0.93 $\pm$ 0.01  & R \\
G308.32-20.23 & &  1208 & 1.00 $\pm$ 0.01 & 0.11 $\pm$ 0.01 & 0.17 $\pm$ 0.01 & 0.45 $\pm$ 0.07 & 0.79 $\pm$ 0.01& 0.27 $\pm$ 0.18 & 0.26 $\pm$ 0.42 & 0.96 $\pm$ 0.01  & R \\
G313.36+61.11 & A1689 & 1348 & 3.08 $\pm$ 0.01 & 0.61 $\pm$ 0.01 & 0.46 $\pm$ 0.01 & 0.63 $\pm$ 0.01 & 0.81 $\pm$ 0.01& 0.02 $\pm$ 0.01 & 0.01 $\pm$ 0.01 & 0.93 $\pm$ 0.01  & R \\
G313.87-17.10 & &  1362 & 2.22 $\pm$ 0.01 & 0.54 $\pm$ 0.01 & 0.39 $\pm$ 0.01 & 0.47 $\pm$ 0.03 & 0.75 $\pm$ 0.01& 0.17 $\pm$ 0.06 & 0.69 $\pm$ 0.21 & 0.98 $\pm$ 0.01  & R \\
G318.13-29.57 & &  1253 & 2.18 $\pm$ 0.03 & 0.85 $\pm$ 0.01 & 0.32 $\pm$ 0.01 & 0.60 $\pm$ 0.13 & 0.76 $\pm$ 0.02& 1.60 $\pm$ 0.98 & 0.02 $\pm$ 1.44 & 0.81 $\pm$ 0.02  & M \\
G321.96-47.97 & A3921 & 1082 & 0.78 $\pm$ 0.01 & 0.40 $\pm$ 0.01 & 0.17 $\pm$ 0.01 & 1.59 $\pm$ 0.04 & 0.63 $\pm$ 0.01& 8.81 $\pm$ 0.29 & 22.70 $\pm$ 1.22 & 0.72 $\pm$ 0.01  & M \\
G324.49-44.97 & RXCJ2218.0-6511 & 974 & 1.33 $\pm$ 0.01 & 0.86 $\pm$ 0.01 & 0.20 $\pm$ 0.01 & 0.61 $\pm$ 0.06 & 0.60 $\pm$ 0.01& 0.39 $\pm$ 0.19 & 0.34 $\pm$ 0.39 & 0.86 $\pm$ 0.01  & R \\
G332.23-46.36 & A3827 & 1236 & 0.98 $\pm$ 0.01 & 0.38 $\pm$ 0.01 & 0.23 $\pm$ 0.01 & 0.58 $\pm$ 0.03 & 0.67 $\pm$ 0.01& 0.36 $\pm$ 0.06 & 0.64 $\pm$ 0.18 & 0.94 $\pm$ 0.01  & R \\
G332.88-19.28 & &  1209 & 1.18 $\pm$ 0.21 & 0.65 $\pm$ 0.01 & 0.22 $\pm$ 0.01 & 1.09 $\pm$ 0.05 & 0.72 $\pm$ 0.01& 0.75 $\pm$ 0.36 & 1.19 $\pm$ 1.01 & 0.95 $\pm$ 0.01  & M \\
G335.59-46.46 & A3822 & 1244 & 0.43 $\pm$ 0.01 & 0.21 $\pm$ 0.01 & 0.11 $\pm$ 0.01 & 0.94 $\pm$ 0.04 & 0.60 $\pm$ 0.01& 0.59 $\pm$ 0.20 & 5.11 $\pm$ 0.91 & 0.94 $\pm$ 0.01  & M \\
G336.59-55.44 & A3911 & 1086 & 0.34 $\pm$ 0.01 & 0.29 $\pm$ 0.01 & 0.09 $\pm$ 0.01 & 1.63 $\pm$ 0.04 & 0.58 $\pm$ 0.01& 0.04 $\pm$ 0.03 & 1.21 $\pm$ 0.27 & 0.86 $\pm$ 0.01  & M \\
G337.09-25.97 & &  922 & 2.40 $\pm$ 0.15 & 0.94 $\pm$ 0.01 & 0.40 $\pm$ 0.01 & 1.09 $\pm$ 0.07 & 0.60 $\pm$ 0.01& 15.40 $\pm$ 1.33 & 61.50 $\pm$ 6.27 & 0.80 $\pm$ 0.01  & M \\
G342.31-34.90 & &  1572 & 0.64 $\pm$ 0.05 & 0.22 $\pm$ 0.01 & 0.16 $\pm$ 0.01 & 0.51 $\pm$ 0.15 & 0.74 $\pm$ 0.02& 1.16 $\pm$ 0.76 & 0.73 $\pm$ 1.66 & 0.90 $\pm$ 0.02  & M \\
G347.18-27.35 & S0821 & 1246 & 0.67 $\pm$ 0.01 & 0.37 $\pm$ 0.01 & 0.08 $\pm$ 0.01 & 2.52 $\pm$ 0.09 & 0.57 $\pm$ 0.01& 0.50 $\pm$ 0.20 & 1.17 $\pm$ 0.77 & 0.96 $\pm$ 0.01  & M \\
G349.46-59.94 & AS1063 & 1446 & 2.91 $\pm$ 0.04 & 0.62 $\pm$ 0.01 & 0.33 $\pm$ 0.01 & 0.88 $\pm$ 0.02 & 0.79 $\pm$ 0.01& 0.03 $\pm$ 0.01 & 0.90 $\pm$ 0.17 & 0.87 $\pm$ 0.01  & R \\
\hline
\end{tabular}
}
\end{table*}

\begin{figure*}[ht!]
\figurenum{J1}
\includegraphics[width=7in]{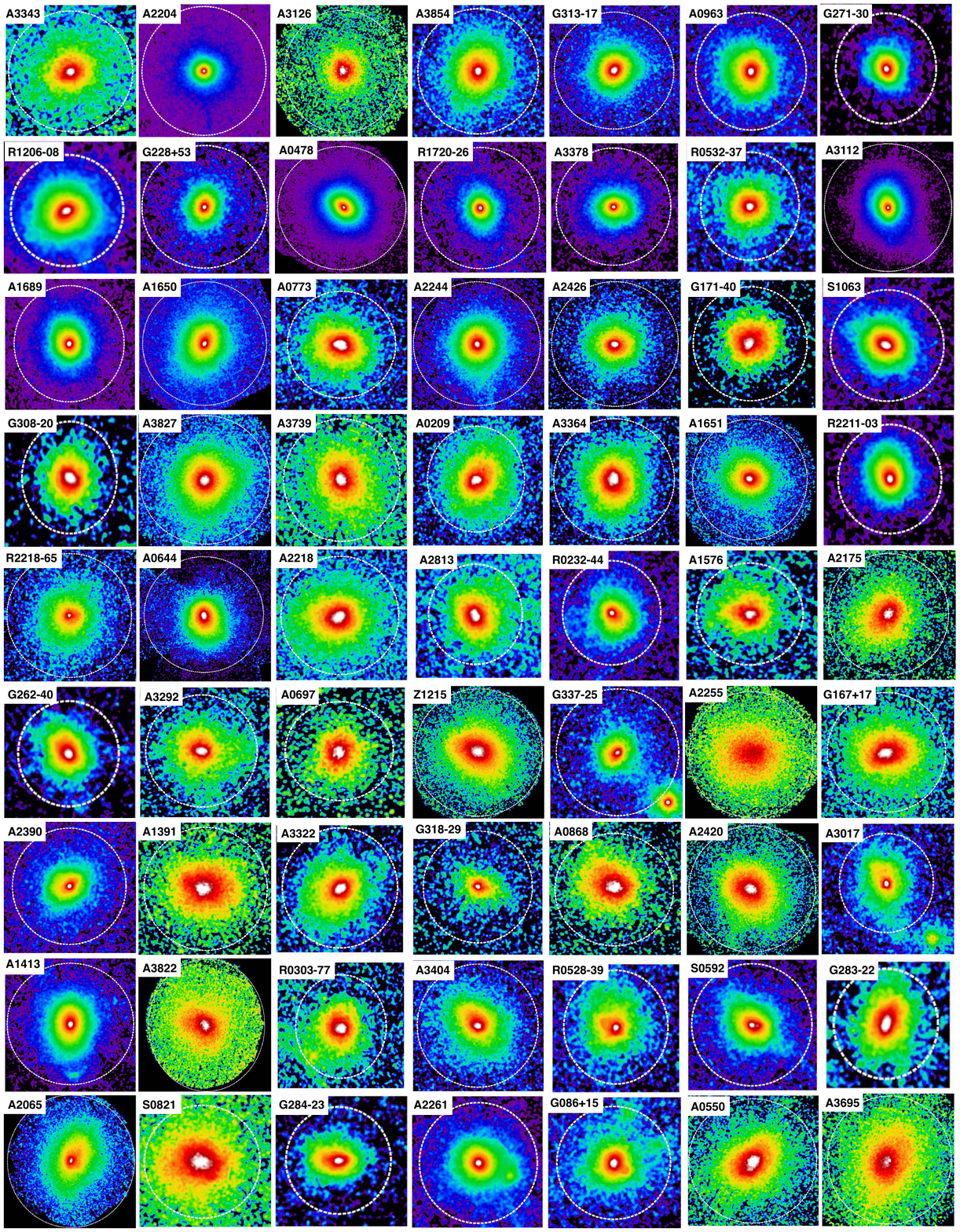}
\caption{X-ray images of the galaxy clusters investigated in this paper, from the most relaxed to the most disturbed accordingly to the visual classification. The white circles indicate the estimated $R_{500}$. }
\end{figure*}

\begin{figure*}[ht!]
\figurenum{J1}
\includegraphics[width=7in]{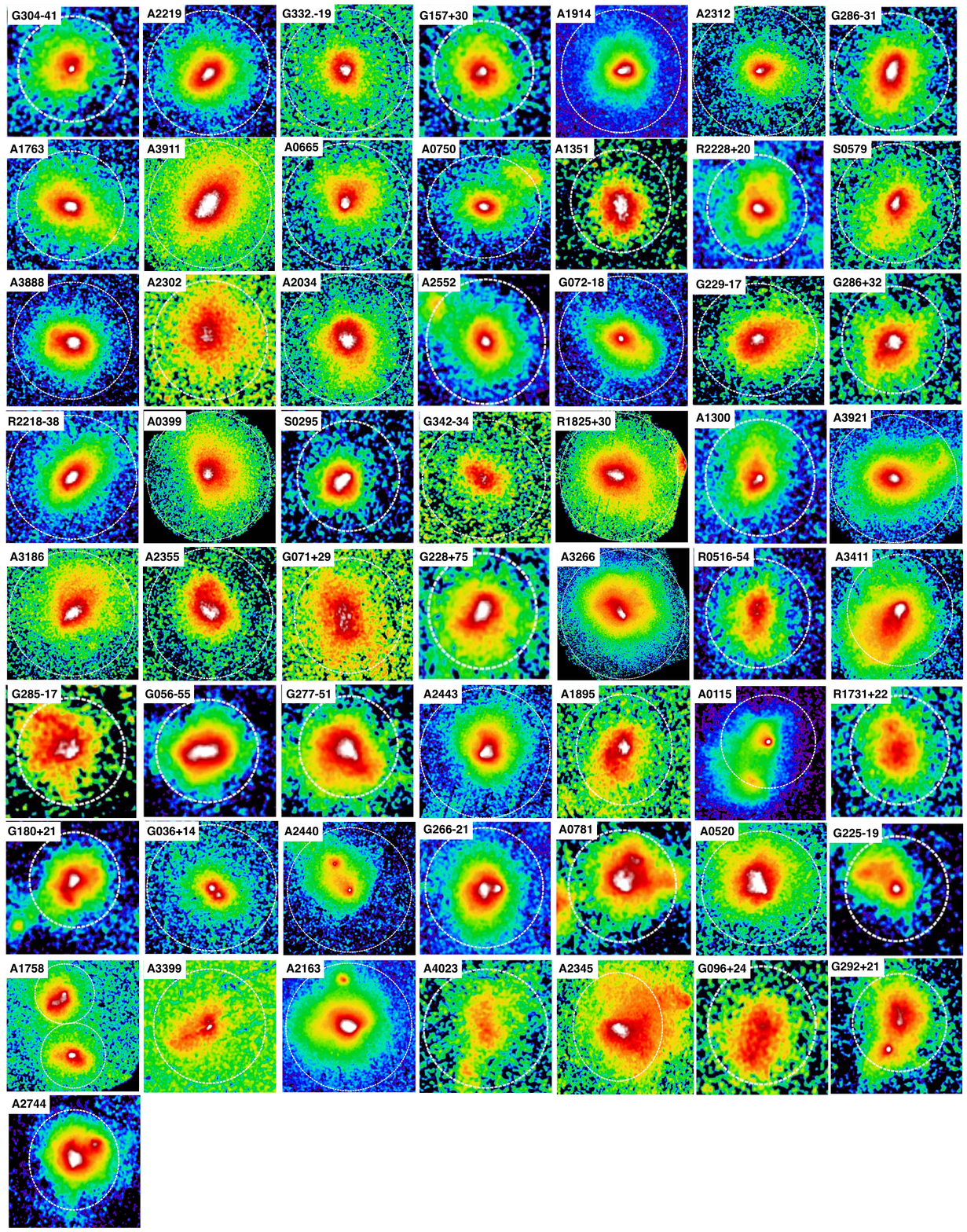}
\caption{continued from the previous page.}
\end{figure*}



\bibliography{biblio}




\listofchanges

\end{document}